\pdfoutput=1
\documentclass[runningheads,a4paper,english]{llncs}

\usepackage[T1]{fontenc}
\usepackage{graphicx}
\usepackage{amssymb}
\usepackage{caption}
\usepackage{booktabs}
\usepackage{yanze}

\newcommand{\lang}{\mathcal{L}_{\mathsf{ART}}}
\newcommand{\pred}{\mathcal{P}}
\newcommand{\pint}{\pred_{\mathit{int}}}
\newcommand{\pext}{\pred_{\mathit{ext}}}
\newcommand{\Hist}{\textsc{Hist}}
\newcommand{\Waker}{\textsc{Waker}}
\newcommand{\Repo}{\textsc{Repo}}      \newcommand{\reactor}{\mathcal{R}}      \newcommand{\lkw}[1]{\texttt{#1}}
\newcommand{\lReady}{\texttt{Ready}}
\newcommand{\lPending}{\texttt{Pending}}
\newcommand{\hist}{\Sigma}                         \newcommand{\Env}{\mathit{Env}}    \newcommand{\env}{\rho}            \newcommand{\Cfg}{\textsc{Config}}    \newcommand{\cfg}{\kappa}          \newcommand{\Global}{\textsc{Global}} \newcommand{\gst}{G}               \newcommand{\wbuf}{\mathcal{Q}}    \newcommand{\Repos}{\textsc{Repos}} \newcommand{\repos}{R}              \newcommand{\pstep}{\longrightarrow_p}

\newcommand{\rstep}{\longmapsto_r}

\newcommand{\precedes}{\prec}
\newcommand{\Obs}{\textsc{Obs}}
\newcommand{\atrue}{\mathsf{tt}}
\newcommand{\afalse}{\mathsf{ff}}
\newcommand{\AbsState}{\textsc{State}}
\newcommand{\tfn}{\mathcal{T}}
\newcommand{\refop}{\mathbin{\triangleright}}

\newcommand{\sWaker}{\widehat{\Waker}}                      \newcommand{\sOb}{\widehat{\textsc{Ob}}}                    \newcommand{\srepos}{\widehat{R}}                           \newcommand{\sloc}{\widehat{L}}                             \newcommand{\cp}{\mathit{cp}}                               \newcommand{\VCR}{\mathit{VC}_{\reactor}}
\newcommand{\VCp}{\mathit{VC}_{\pi}}

\newcommand{\reactoronly}[1]{\colorbox{cyan!15}{\ensuremath{#1}}}

\DeclareRobustCommand{\asout}[1]{}

\newcommand*{\figref}[1]{Fig.~\ref{fig:#1}}
\newcommand*{\secref}[1]{Sec.~\ref{sec:#1}}
\newcommand*{\defref}[1]{Def.~\ref{def:#1}}
\newcommand*{\appref}[1]{App.~\ref{app:#1}}

\newcommand{\eg}{{{e.g.,\@}}} 
\newcommand{\ie}{{{i.e.,\@}}}  \newcommand{\etal}{{{et~al.\@}}}

\makeatletter

\def\operator#1{\@ifnextchar\bgroup {\operatorarg{\ensuremath{#1}}}{\ensuremath{#1}}}
\def\operatorarg#1#2{{#1}{\ensuremath{(#2)}}}

\def\spoperator#1#2{\@ifnextchar\bgroup{\spoperatorarg{\ensuremath{#1}}{\ensuremath{#2}}}{\ensuremath{#1}}}
\def\spoperatorarg#1#2#3{\ensuremath{#1#2#3}}

\makeatother

\newcommand{\task}{\operator{\ensuremath{\tau}}}
\newcommand{\public}{public\@} \newcommand{\Ob}{\operator{\textit{Ob}}}
 
\spnewtheorem{tcomp}{Trusted Component}{\bfseries}{\itshape}

\begin{document}
\title{Modular Responsiveness Verification of Rust Async Runtimes}
\author{Yanze Li \and Ivan Beschastnikh \and Alexander J. Summers}
\institute{University of British Columbia, Canada}

\maketitle

\begin{abstract}
Asynchronous (async) programming is a popular paradigm for managing concurrency. Languages that provide async support typically have a runtime to manage asynchronous executions. These runtimes are critical infrastructure, yet verifying them has received little attention. One reason is that a property that users care most about from an async runtime is a liveness property: tasks submitted to the runtime eventually make progress. Verifying liveness is challenging for libraries that are both concurrent and highly optimized.

We present a lightweight and modular proof technique for verifying eventual progression guarantees for Rust async runtimes. We describe this technique in the context of a simple language based on Rust and Rust's async model. We then realize this proof technique as a set of static analyses for Rust and use these to verify eventual progression of several key components of multiple Rust async runtime implementations.

\end{abstract}

\section{Introduction}
\label{sec:intro}

To take advantage of multi-core CPU resources, modern programs make extensive use of asynchronous programming, which relieves developers from needing to manage concurrency themselves.
Asynchronous programming is usually supported by an \emph{async runtime} and such runtimes exist for most languages, including Java~\cite{oracle_completablefuture,guava_listenablefuture}, JavaScript~\cite{mdn_async_await}, Python~\cite{python_asyncio}, and Rust~\cite{smol_docs,tokio_docs}.

An async runtime takes care of managing a graph of inter-dependent user-defined asynchronous tasks, some of which may be waiting on external events, and others that may be waiting on the result of other tasks.
Async runtimes vary in complexity, but even a simple runtime must keep track of tasks, wait and observe relevant events, and notify tasks to continue executing. These runtimes and their correctness from a user's perspective are the focus of this paper.

A critical \emph{responsiveness property} of async runtimes that users depend on is \emph{eventual progression}: assuming correct user code and a correct underlying operating system, every user-level async task managed by the runtime should eventually get to make progress (see Section~\ref{sec:overview}).

The eventual progression property is the cornerstone of any async runtime, but is also challenging to verify.
This is because a typical async runtime manages a collection of async tasks, and these tasks may depend on a variety of external events.
Async runtimes are also usually concurrent and contain multiple components that communicate with one another.
In addition to this inherent complexity of async runtimes, languages such as Rust and OCaml do \emph{not} ship an async runtime as part of the languages but only provide an interface for implementing async runtimes; users can use different runtime implementations simply by importing different async libraries.
This means that any formal techniques aimed at async runtimes for such languages must be able to handle a diversity of alternative implementations in practice. 

\textbf{In this paper, we present a lightweight verification technique that allows programmers to verify eventual progression for practical and diverse async runtime implementations.}
To the best of our knowledge, this is the first work to verify liveness guarantees of existing async runtime implementations.

Rust async runtimes consist of three major parts:
\begin{enumerate*}
    \item async tasks, which include primitive async tasks provided by async runtimes (called \emph{pollables}) and user-defined async functions using pollables,
    \item Rust's async/await syntax and compiler support, which compile async functions to state-machine representations, and
    \item \emph{reactors}, which manage running async tasks by monitoring signals, waking up tasks when signals are observed, and repeating until all tasks complete.
\end{enumerate*}
Assuming correctness of the Rust compiler, eventual progression boils down to the behaviors of the sets of runtime-provided pollables and reactors.

Our work relies on three key insights.
First, we adopt a \emph{task-centric view} and find that it is sufficient to consider verification of eventual progression for a single arbitrary task, instead of reasoning about a collection of tasks.
Second, we use runtime component boundaries for a natural and modular decomposition of our verification goal.
Finally, we observe that runtime components make use of implicit and, most importantly, \emph{monotonic} events to determine how to deal with a task at any particular point in time. These events are usually internal to the runtime and are key to establishing eventual progression.

We formalize these three insights in a core language $\lang$ with an operational semantics for the implementation of pollables and reactors. We then define static analyses for $\lang$ whose verification conditions soundly imply eventual progression of the runtime.
Grounded in our formalism, we design a lightweight specification language and verification technique for Rust called \tool{}.
Developers of async runtimes can specify the necessary conditions for pollables to be responsive, and annotate their implementation in terms of observations on events.
We use \tool{} to verify eventual progression of pollable and reactor components in two commonly used Rust async runtime libraries, smol~\cite{smol_docs} and futures-rs~\cite{futures_docs}, and tokio~\cite{tokio_docs}, the most widely used production runtime.
We show that our model and analyses handle a variety of real-world components with two modest extensions.

Our evaluation shows that pollable implementations can be verified with an average specification-to-code ratio of less than 30\%, and the verification can prevent manually-seeded responsiveness bugs of various kinds.

In summary, the main contributions of our work are as follows:
\begin{enumerate}
    \item We develop an abstract model of Rust async runtimes, via our study of real-world implementations. Using our model, we observe that eventual progression for async runtimes fundamentally relies on the implementations of two kinds of critical components: \emph{pollables} and \emph{reactors}, which are programmed in a highly-stylized way.
    We formalize a core language $\lang$ capturing this programming paradigm for modeling pollable and reactor implementations.
    \item We formally define the key responsiveness property of interest, \emph{eventual progression}, in terms of an operational semantics for $\lang$ that uses a novel notion of \emph{obligations}.
   We develop modular static analysis techniques that track how each obligation is eventually discharged, and show their soundness and that this implies eventual progression.
\item We implement a prototype tool \tool{} in Rust, which allows annotating Rust pollable and reactor implementations to obtain models in $\lang$, and analyzes these models. We evaluate \tool{} on three Rust async runtimes and successfully verify eventual progression of various pollable/reactor implementations with low specification effort. We also show that \tool{} rejects responsiveness bugs that can arise in this challenging code.
\end{enumerate}

 \section{Background}
\label{sec:bg}

Async programming, as a concurrency programming model, provides a user-level abstraction (called \emph{async tasks}) for writing concurrent programs.
Compared to directly using OS threads, async tasks are entirely managed by a runtime system in user space; this lets programs run a larger number of computations concurrently on a small number of OS threads (even a single thread).

\subsection{Async Rust}
\label{sec:bg-async-rust}

Rust provides \rcode{async/await} syntax for writing async programs in a style similar to sequential ones.
One can define an \textit{async function} by adding the \rcode{async} keyword before a normal function definition, such as the identity function below:
\begin{rust}
async fn id<T>(x: T) -> T { return x; }
\end{rust}
When the async function \rcode{id} is called, instead of returning a value of declared type \rcode{T}, a \emph{future} of type \rcode{Future<T>} is returned.
This future represents an \textit{async computation} that, once run, should eventually resolve to a value of type \rcode{T}.

Within async functions, one can attempt to resolve async computations by \emph{awaiting} on them.
The \rcode{await} expression in the example below executes the \rcode{id} function and immediately resolves it to the string literal.

\begin{rust}
async fn hello_world() {
    let s = id("hello world").await;
    println!(s);
}\end{rust}
Await-points define where an async computation might need to be paused.
This does not happen in our \rcode{hello\_world} example for two reasons:
\begin{enumerate}
    \item The function only awaits async computations, which then run immediately.
    \item There are no other computations that can be run concurrently.
\end{enumerate}
These missing pieces are provided by combining \rcode{async/await} with async runtime-provided \textit{pollables}: a collection of primitive async computations such as async I/O handlers, timers, and channels.
Their implementations are provided by the runtime directly (and not as user-level async functions); each call to a pollable will return directly rather than be decomposed into smaller sub-computations.

The common usage of \rcode{await} reflects the design that all runtime-provided pollables implement the same \rcode{Future} trait; this syntax is compiled into calls to the \rcode{poll} method declared as follows:
\begin{rust}
pub trait Future<T> { fn poll(&mut self, w: Waker) -> Poll<T>; }\end{rust}
The \rcode{poll} method starts or resumes an async computation, executes until the computation yields, and returns the state of the async computation: \rcode{Pending} if the computation yields, and \rcode{Ready} if the computation completes\footnote{The method name \rcode{poll} may be somewhat misleading since \rcode{poll} does not simply query the state of async computations but also executes them as far as possible.}:
\begin{rust}
pub enum Poll<T> { Ready(T), Pending }
\end{rust}
One might assume that user-defined async computations are expected to busy-poll their dependent pollables.
Instead, an async computation that awaits another which is pending will itself be considered pending, and the \emph{runtime} is given the responsibility to track pending tasks and resume them when appropriate.

The Rust compiler converts each async function into a state-machine: each await-point in an async function, along with its start and end, makes up the states.
The state machine is implemented as the (generated) \rcode{poll} method for the \rcode{Future} object returned from the async function call.
Correspondingly, all \rcode{await} expressions are translated to calls on the \rcode{poll} method of the awaited-on task.
When such calls return with a \rcode{Ready} result, the state machine for the \rcode{await}ing task will advance; otherwise, a \rcode{Pending} result will be propagated upwards.
This enables programmers to compose larger async programs from smaller ones, starting from pollables provided by the chosen runtime.

To enable concurrency when a user task is paused, a runtime must have alternative tasks to schedule.
Runtime implementations provide an interface, typically named \rcode{spawn}, for registering potentially-concurrent tasks:
\begin{rust}
async fn concurrent_prints() {
    let t1 = spawn(print_after_3s());
    let t2 = spawn(print_after_3s());
    t1.await;
    t2.await;
}\end{rust}
When task \rcode{t1} starts waiting for its timer, the runtime can concurrently execute task \rcode{t2}.
It is the runtime's responsibility to guarantee that, sometime after the waited-for time expires, the \rcode{poll} method of the timer will be called again, \emph{and as a result}, the corresponding user task will resume executing.

In summary, async programming in Rust relies on the built-in \rcode{async/await} support of the language, but more importantly on an async runtime providing correctly implemented interfaces such as \rcode{Timer} and \rcode{spawn}.
Since Rust does not ship with a dedicated async runtime, different async programs can use different runtime implementations.
It is therefore important that different async runtimes live up to the same set of specifications.

\subsection{Rust Async Runtimes}
\label{sec:bg-rust-async-rt}

\begin{figure}[t]
    \centering
    \includegraphics[width=\textwidth]{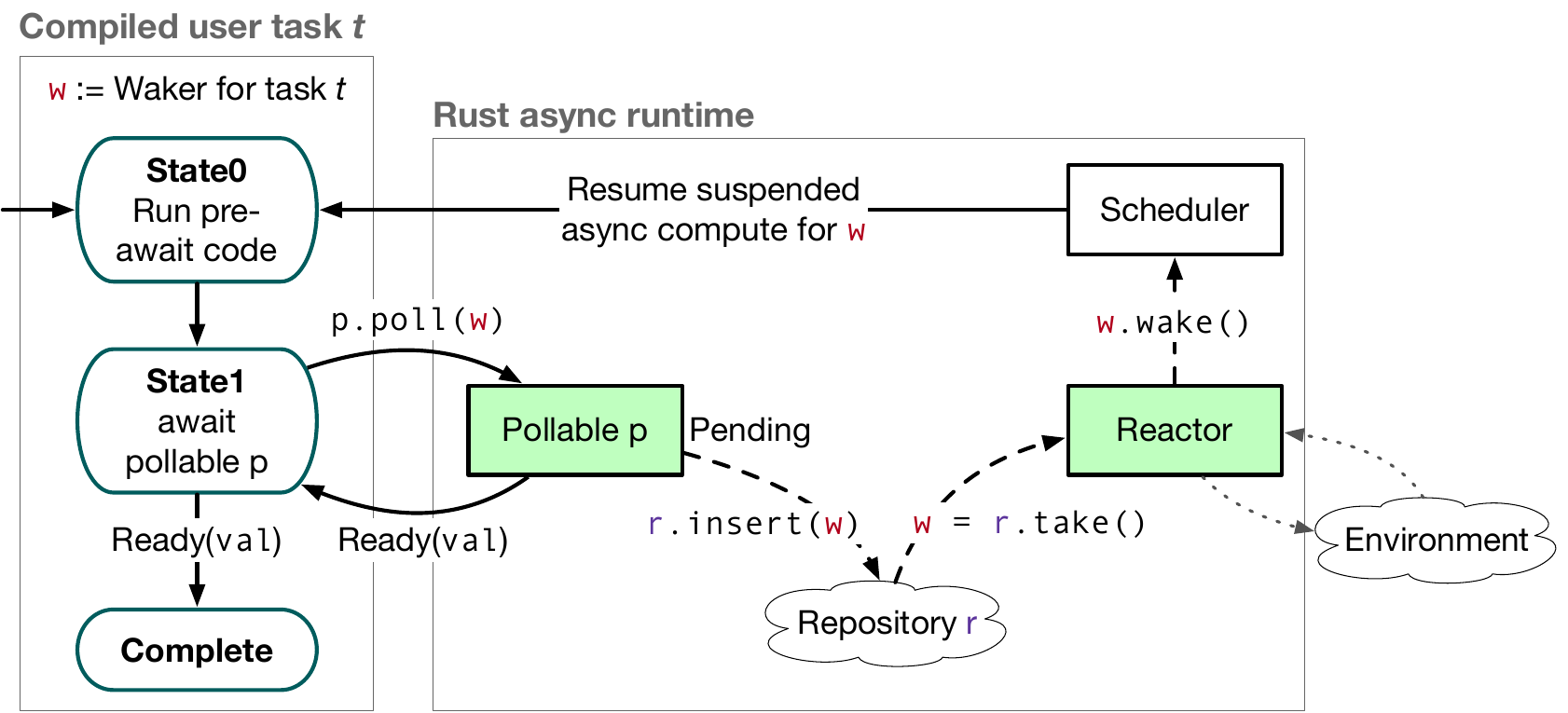}
    \caption{Control flow between Pollable/Reactor/Waker/Scheduler. Each Pollable returns \rcode{Ready} or \rcode{Pending}; the Reactor reacts to the environment. Repositories are a concept we introduce in this paper to abstract over how wakers are passed to and stored by reactors.}
    \label{fig:flow}
\end{figure}

The progression of user-defined async tasks relies on an async runtime (Figure~\ref{fig:flow}) whenever those tasks are paused.
A critical responsibility of the runtime is that any paused task must eventually have its \rcode{poll} method called, attempting to resume its execution.

\textbf{Wakers.}
Rust async runtimes encapsulate a pending-resumption task in a \rcode{Waker} type.
Each \rcode{Waker} is a handle on the top-level user task whose poll chain currently (transitively) awaits a pollable. Calling \rcode{wake} on it notifies a \textit{scheduler} to re-invoke that top-level task's \rcode{poll}. The compiler-generated state machine records the await-point at which the task was suspended, so the re-entered \rcode{poll} resumes there and re-invokes \rcode{poll} on the same awaited pollable that previously returned \rcode{Pending}.
The only possibility for these tasks to be paused is if a runtime pollable returns \rcode{Pending}.
It is the responsibility of the runtime implementation to register the corresponding \rcode{Waker} with the runtime, along with information to allow the runtime to determine what event the task is waiting for.

\textbf{Reactors.}
A dedicated runtime component (typically called a \textit{Reactor}) is used to manage all currently-paused tasks, storing references to their corresponding \rcode{Waker} instances.
The Reactor is run continuously: its responsibility is to periodically check the \textit{conditions} that each paused task is waiting on (e.g., whether timers have expired, or I/O events occurred).
Rather than directly calling \rcode{poll} on these tasks, which would interrupt the Reactor's execution, the Reactor calls \rcode{wake} to register the task with a scheduler.
Assuming a fair scheduling algorithm, the scheduler guarantees that each of these tasks will ultimately have its \rcode{poll} function called again.

In this paper, we leverage the abstractions provided by this idiomatic design (which is shared across all Rust async runtimes we are aware of), and assume in particular a high-level specification for wakers (and their scheduling) without verifying their implementations or the fairness of the scheduling algorithm. Instead, we focus on
\begin{enumerate*}[label=(\arabic*)]
    \item the implementations of pollables, which form the key building blocks and interfaces for building user-level programs, and
    \item the implementations of Reactors, which are responsible for ensuring that these pollables, and therefore the dependent user tasks, get called again as needed.
\end{enumerate*}

 \section{Overview}
\label{sec:overview}

We introduce our approach to verifying eventual progression of Rust async runtimes through a representative example: a \rcode{Timer} pollable paired with a Reactor that manages a collection of timers.
We then state the eventual progression property and the trusted components on which it relies, and informally describe the per-component conditions our verification technique enforces.
The model and the analyses are made formal in Sections~\ref{sec:model}~and~\ref{sec:verification}.

\subsection{The Timer Example}
\label{sec:overview-timer}

A \rcode{Timer} is a runtime-provided pollable that pauses the calling async task until a specified time has elapsed.
Figure~\ref{fig:timer-poll-impl} shows the relevant data definitions and the \rcode{poll} method of \rcode{Timer}.
A timer carries the target time \rcode{instant} and a reference to the Reactor responsible for waking it up.
The \rcode{poll} method first compares \rcode{instant} against the current time \rcode{now()}.
If the time has already passed, \rcode{poll} returns \rcode{Ready} and the awaiting task proceeds.
Otherwise,
\rcode{poll} sends the waker and \rcode{instant} to the Reactor via \rcode{transfer},
and returns \rcode{Pending}. The awaiting task is paused, and the Reactor takes over the responsibility to call \rcode{poll} again when the target time has passed.

\begin{figure}[tbp]
    \centering
    \begin{rust}
struct TimerData { instant: Instant, waker: Waker }
struct Timer { instant: Instant, reactor: &Reactor }

// w is a Waker for this pollable (guaranteed by Rust compiler)
fn poll(self, w: Waker) -> Poll<()> {
    if self.instant <= now() {
        return Poll::Ready(());
    } else {
        self.reactor.transfer(
            TimerData { instant: self.instant, waker: w });
        return Poll::Pending;
    }
}\end{rust}
    \caption{A \rcode{Timer} pollable. \rcode{transfer} inserts a \rcode{TimerData} (the waker together with the target time) into the Reactor's repository. We elide Rust details concerning references and the \rcode{Future} trait.}
    \label{fig:timer-poll-impl}
\end{figure}

A typical Reactor that manages a collection of timers is shown in Figure~\ref{fig:reactor-with-repo}.
The Reactor maintains all currently-paused timers in the \rcode{timers} vector; \rcode{transfer} (elided) inserts a fresh \rcode{TimerData} into this vector under appropriate synchronization.
The \rcode{react} method runs an infinite loop: at each iteration, it partitions the vector based on whether each timer's target time has passed, wakes every timer whose time has passed, and retains the others for the next iteration.
The Reactor cannot terminate, because new responsibilities to wake timers may be transferred to it at any future point.

\begin{figure}[tbp]
    \centering
    \begin{rust}
struct Reactor { timers: Vec<TimerData> }

fn react(&self) {
    loop {
        let (ready, pending) = self.timers.partition(
            |t: TimerData| t.instant <= now()
        );
        for t in ready { t.waker.wake(); }
        self.timers.append(pending);
    }
}\end{rust}
    \caption{A Reactor that manages a collection of timers. The \rcode{react} method is spawned as a background thread.}
    \label{fig:reactor-with-repo}
\end{figure}

This pair illustrates the three reasoning concepts our verification builds on:

\textbf{Monotonic completion conditions.}
A timer's completion condition \rcode{self.instant <= now()} is a predicate on the external environment that is \emph{monotonic}: once true, it remains true.
Monotonicity is what allows \rcode{poll} (and the Reactor) to reliably decide completion based on an observation of the condition.
Pollables other than timers (e.g., I/O handlers, channels) condition on different monotonic predicates, but the principle is the same.

\textbf{Responsibility transfer between components.}
Initially the called pollable is responsible for its own completion: if its completion condition holds when \rcode{poll} is called, \rcode{poll} returns \rcode{Ready} directly.
If not, the pollable hands the responsibility to its Reactor via \rcode{transfer}, then returns \rcode{Pending}.
The Reactor periodically checks the completion condition and, once it holds, calls \rcode{wake} on the corresponding waker; the wakened task is then re-scheduled, and the responsibility is finally discharged by a \rcode{poll} call returning \rcode{Ready}.

\subsection{Eventual Progression: the Core Responsiveness Property}
\label{sec:overview-ep}

Informally, the property we want of an async runtime is: \emph{whenever a user task is paused waiting on a pollable whose completion condition will eventually hold, that task is eventually resumed.}
We refer to this property as \emph{eventual progression}.
We make this precise in Section~\ref{sec:model-properties}.

Several components participate in delivering eventual progression: the user task itself, the Rust compiler, the external environment, the scheduler, and the runtime-provided pollables and Reactors.
A bug anywhere along this chain can violate the property; we cannot ask any single component to bear the entire burden.
Our verification focuses on the pollable and Reactor implementations, treating the other components as \emph{trusted}.

\begin{tcomp}[The Rust Compiler]
The Rust compiler preserves the semantics of async functions during the compilation of these functions to their state machine representations (Section~\ref{sec:bg-async-rust}).
\end{tcomp}

\begin{tcomp}[Observations of the External Environment]
All APIs for polling and extracting data from the environment external to the Rust program (OS, devices, etc.) behave correctly: signaling events and providing data when expected.
\end{tcomp}

\begin{tcomp}[Fair Scheduling of Rust Async Tasks]
The scheduler components of the Rust async runtime schedule tasks passed to them fairly.
\end{tcomp}

The first two trusted components reflect well-established compilation and OS infrastructure; the third reflects a well-studied problem largely independent of the async focus of our work.
Under these assumptions, the responsibility for eventual progression reduces to the interaction between pollables and Reactors: it is their collaboration that translates ``the completion condition holds'' into ``the awaiting task's \rcode{poll} is called and returns \rcode{Ready}.''

\subsection{Per-Component Conditions}
\label{sec:overview-conditions}

We decompose eventual progression into two component-level conditions, one for pollables and one for Reactors.
Both are stated here informally; we make them precise as verification conditions in Section~\ref{sec:verification}.

A pollable's responsibility is captured by three conditions on its \rcode{poll} method:

\begin{enumerate}
    \item[(P1)] \emph{Completion faithfulness.} If \rcode{poll} returns \rcode{Ready}, the completion condition has held.
    \item[(P2)] \emph{Completion convergence.} If the completion condition holds when \rcode{poll} is called, \rcode{poll} returns \rcode{Ready}.
    \item[(P3)] \emph{Responsibility transfer.} If \rcode{poll} returns \rcode{Pending}, the pollable has transferred its waker to a Reactor along with its completion condition.
\end{enumerate}

A Reactor's responsibility is captured by two iteration-level conditions on the body of its main loop:

\begin{enumerate}
    \item[(R1)] \emph{Iteration progress.} If a transferred waker's completion condition holds at the start of an iteration, the Reactor calls \rcode{wake} on it within that iteration.
    \item[(R2)] \emph{Iteration preservation.} A transferred waker that is not woken in this iteration is retained in the Reactor's repository for a future iteration.
\end{enumerate}

We illustrate these conditions using the \rcode{Timer}/Reactor pair from Section~\ref{sec:overview-timer}.
The completion condition is \rcode{self.instant <= now()}.
\rcode{Timer}'s \rcode{poll} returns \rcode{Ready} only on the path where \rcode{self.instant <= now()} is observed, so (P1) holds.
If \rcode{self.instant <= now()} holds when \rcode{poll} is called, the \rcode{if}-test takes the \rcode{Ready} branch, so (P2) holds.
The \rcode{Pending} path is preceded by \rcode{transfer}, which inserts \rcode{TimerData} into the Reactor's repository; this is the responsibility transfer (P3).
For the Reactor: \rcode{partition} separates timers whose target time has passed from those that have not; \rcode{wake} is then called on every timer in the first group, satisfying (R1); \rcode{append} retains the rest, satisfying (R2).

Crucially, (P1)--(P3) and (R1)--(R2) can be checked per-component, in isolation.
This modularity is provided by our task-centric view: even though a Reactor concretely manages a collection of timers, the per-iteration conditions can be discharged by reasoning about a single arbitrary timer.
Sections~\ref{sec:model} and~\ref{sec:verification} formalize this insight by defining a core language $\lang$ for pollable and Reactor implementations, an abstract domain that tracks observations of monotonic completion conditions, and verification conditions that imply (P1)--(P3) and (R1)--(R2).

 \section{A Model of Rust Async Runtimes}
\label{sec:model}

We model Rust async runtime components in a core language $\lang$ (\underbar{L}anguage for \underbar{A}sync \underbar{R}un\underbar{t}ime).
$\lang$ is designed to capture the high-level \emph{architecture} of Rust async runtimes and the programming paradigm for pollable and reactor reasoning rather than modeling the programming language Rust.

Importantly, $\lang$ builds in two primitive notions that are central to verifying eventual progression: \emph{monotonic predicates} (typically to model completion predicates for pollables) and \emph{obligations} (used to abstractly represent suspended user tasks waiting on a runtime-provided pollable).

To reason about a collection of obligations in a reactor, we build in the notion of a \emph{repository} as an abstraction over concrete ways a reactor may store wakers.
A repository abstracts over a concrete data type such as a Rust \rcode{Vec}, and $\lang$ statements abstract over concrete functions how wakers are passed to (\eg{} \rcode{transfer} in \figref{timer-poll-impl}), stored in, and removed from these data types.

$\lang$ is intentionally minimal, but as we will show in Sections~\ref{sec:verification} and~\ref{sec:impl-eval}, $\lang$ is sufficient to capture the design idioms shared by real Rust async runtime components with few extensions.

\subsection{The Language $\lang$}
\label{sec:model-language}

$\lang$ is parametrized over an infinite set of \emph{monotonic predicate symbols} $\pred$ (hereafter, simply \emph{predicates}), partitioned into two disjoint (infinite) sets of \emph{internal} and \emph{external} predicates: $\pred = \pint \uplus \pext$.
A monotonic predicate $p \in \pred$ represents a \emph{stable-once-established} condition: once $p$ holds, it holds permanently in the future.
\emph{Internal} predicates represent conditions that can be satisfied by program actions (\eg{} setting a flag from \rcode{false} to \rcode{true}); \emph{external} predicates model conditions which can only be updated by the environment, \eg{} external signals such as a timer expiring or an I/O event completing.

In real code, the truth of monotonic predicates can typically be tested via suitable boolean Rust expressions whose value changes monotonically; for example, a timer's completion condition can be checked via \rcode{self.instant <= now()}. We use the predicate symbols in our language to abstract over these concrete conditions, which are required only to satisfy monotonicity.

We assume an infinite set of wakers, $\Waker$, and a lookup function $\tau : \Waker \rightharpoonup \Stmt \times \Env$;
for a waker $w$, $\task{w}$ denotes the (compiled, suspended) user task the waker is able to reschedule (\secref{bg-rust-async-rt}).
As will be explained in \secref{model-opsem}, such suspended tasks are represented concretely by pairs of \emph{a pollable implementation} that the user task is currently waiting on and its local environment; in the real Rust code this is handled with compiler-generated closures.

\begin{figure}[tbp]
    \centering
    \begin{align*}
    \text{Values}     \quad & v  ::= ()
                              \;\mid\; \lReady(v)
                              \;\mid\; \lPending
                              \;\mid\; w \quad (w \in \Waker)\\
    \text{Statements} \quad & s  ::= s_1 \,;\, s_2
                              \;\mid\; \lkw{skip}
                              \;\mid\; \lkw{return}\; v
                              \;\mid\; \lkw{wake}(w) \\
                            & \phantom{s ::=}\;\mid\; \lkw{trigger}(p) \;\; (p \in \pint) \\
                            & \phantom{s ::=}\;\mid\; \lkw{if}\; p\; \lkw{then}\; s_1\; \lkw{else}\; s_2 \quad (p \in \pred) \\
                            & \phantom{s ::=}\;\mid\; \lkw{insert}(r, w, p) \;\; (r \in \Repo,\; p \in \pred) \\
                            & \phantom{s ::=}\;\mid\; \reactoronly{ \lkw{take}(r, w) } \;\; (r \in \Repo)
                              \;\mid\; \reactoronly{ \lkw{for}\; (w, p)\; \lkw{in}\; r\; \{\, s \,\} } \\
                            & \phantom{s ::=}\;\mid\; \reactoronly{ \lkw{loop}\; s } \\
    \text{Pollables} \quad & \pi  ::= \lambda (w, p).\, s \quad (\mathit{wfp}(s))\\
    \text{Reactors}   \quad & \reactor ::= \lkw{reactor}\; \{\,
                                \lkw{shared}\; \vv{r_s : (T_{r_s}, \cp_{r_s})}\;
                                \lkw{in}\; (\lkw{loop}\; s)
                              \,\} \quad (\mathit{wfr}(s))
    \end{align*}
    \caption{Syntax of $\lang$, parametrized over monotonic predicates $\pred = \pint \uplus \pext$. A pollable is a statement that returns on every path; a reactor declares a set of \public{} repositories and wraps its body in an infinite loop. We $\reactoronly{\text{highlight}}$ statements that are reactor-only.}
    \label{fig:lang-syntax}
\end{figure}

Figure~\ref{fig:lang-syntax} defines the syntax of $\lang$.
Values include the unit, the pollable return values $\lReady(v)$ and $\lPending$, and wakers.
Statement $\lkw{trigger}(p)$ sets the value of an internal predicate $p$ to true (it is a no-op if $p$ was already true).
$\lkw{insert}(r, w, p)$ sends the waker $w$ to the \public{} repository $r$ together with the completion predicate $p$ that must be waited for.
Conceptually, this transfers the obligation for \emph{progress} of $\task{w}$ to the reactor $r$ (which must now monitor $p$ and wake $w$ when satisfied).
$\lkw{take}(r, w)$ removes the entry for $w$ from $r$ (\eg{} because it will be woken up).
$\lkw{wake}(w)$ invokes waker $w$, signaling the scheduler to re-schedule the user task $\task{w}$, which will in turn re-poll \emph{the corresponding pollable}.
$\lkw{for}\;(w, p)\;\lkw{in}\;r\;\{\,s\,\}$ iterates the body $s$ over every $(w, p)$ pair in $r$.
$\lkw{loop}\;s$ repeatedly runs $s$ (indefinitely).
$\lkw{return}\;v$ terminates a pollable with result $v$ (typically $\lReady(())$ or $\lPending$).

We impose \emph{well-formedness conditions} (formally defined in Appendix~\ref{app:wf}) on the $\lang$ statements representing pollable and reactor implementations.
A pollable $\pi$ is a statement that satisfies its well-formedness predicate $\mathit{wfp}$ when $\pi$ contains no $\lkw{loop}$ or reactor-only constructs, and every execution path through $\pi$ contains a $\lkw{return}$ statement.

Reactors $\reactor$ are first-class in $\lang$, and each owns (\ie{} no overlap with other reactors) a non-empty list of \emph{\public{} repositories}.
Each \public{} repository $r$ is declared along with a pair $(T_r, \cp_r)$, in which $T_r$ is the type of pollable whose wakers may be stored in $r$ and $\cp_r : T_r \to \pred$ is the \emph{completion predicate scheme} mapping each pollable instance to its completion predicate.
A reactor body needs to satisfy its well-formedness predicate $\mathit{wfr}$, which requires that the body never $\lkw{return}$s and contains no $\lkw{loop}$ other than the (required) outermost one; this reflects the fact that reactors are intended to monitor the environment indefinitely and never return.

For simplicity, we assume each reactor has at most one \public{} repository per type of pollable and that well-formed programs respect the repository type-checking constraints.
We also assume the convention that $\lkw{insert}$, $\lkw{take}$, and $\lkw{for}$ in a reactor body access only repositories declared by that reactor, and pollables always communicate with a dedicated, running reactor.

In real code, rather than being associated with completely independent predicates, two pollables of the same type will always depend on \emph{similar} conditions, but may differ in instance-specific parameters: for example, two timers both wait on $t \leq \lkw{now()}$, but with different deadlines $t$. In our language, we simply represent these as different predicates under the same scheme, as we abstract over the form of these concrete boolean expressions.

\subsection{Operational Semantics}
\label{sec:model-opsem}
We define a small-step operational semantics for $\lang$; aside from the language itself the main design novelty is that our runtime states explicitly track the known status of monotonic predicates, as well as the status of wakers for suspended user tasks. We begin by defining rules for statement execution (only), and then show how to lift these to accommodate rules that model interleavings with \eg{} events from the external environment.

A runtime state is a pair $(\env, \gst) \in \Cfg \triangleq \Env \times \Global$, separating local states from shared global states:
\begin{itemize}
    \item $\env \in \Env$ is the \emph{local environment} for pollables or reactors, mapping local variables to values;
    \item $\gst = (\hist, \wbuf, \repos) \in \Global \triangleq \Hist \times \wp(\Waker) \times \Repos$ is the \emph{global state}, where
    \begin{itemize}
        \item $\hist \in \Hist \triangleq \wp(\pred)$ is the \emph{monotonic history}, recording the set of monotonic predicates true so far;
        \item $\wbuf \in \wp(\Waker)$ is the \emph{rescheduling queue}, containing the wakers that have been woken up but waiting for the scheduler to reinvoke the \rcode{poll} method of the corresponding pollable;
        \item $\repos \in \Repos \triangleq \Repo \to \wp(\Waker \times \pred)$ is the \emph{repository map}, mapping each repository to the set of waker-predicate pairs stored in it. We assume all waker-predicate pairs in a repository $r$ respect the repository typing, i.e., every $(w, p) \in \repos(r)$ has $p \in \mathit{img}(\cp_r)$, where $\mathit{img}$ denotes the image of the predicate scheme $\cp_r$.
    \end{itemize}
\end{itemize}

\begin{figure}[tbp]
    \raggedright
    \boxed{
    s \vbar \cfg \pstep s' \vbar \cfg'
    }
    \begin{mathpar}
    \inference{
    p \in \pint
    }{
    \lkw{trigger}(p) \vbar (\env, (\hist, \wbuf, \repos)) \pstep \lkw{skip} \vbar (\env, (\hist \cup \{p\}, \wbuf, \repos))
    }[(\text{trigger})]
    \and
    \inference{
    }{
    \lkw{wake}(w) \vbar (\env, (\hist, \wbuf, \repos)) \pstep \lkw{skip} \vbar (\env, (\hist, \wbuf \cup \{w\}, \repos))
    }[(\text{wake})]
    \and
    \inference{
    \repos' = \repos[r \mapsto \repos(r) \cup \{(w, p)\}]
    }{
    \lkw{insert}(r, w, p) \vbar (\env, (\hist, \wbuf, \repos)) \pstep \lkw{skip} \vbar (\env, (\hist, \wbuf, \repos'))
    }[(\text{insert})]
    \and
    \inference{
    (w, p) \in \repos(r) \\
    \repos' = \repos[r \mapsto \repos(r) \setminus \{(w, p)\}]
    }{
    \lkw{take}(r, w) \vbar (\env, (\hist, \wbuf, \repos)) \pstep \lkw{skip} \vbar (\env, (\hist, \wbuf, \repos'))
    }[(\text{take})]
    \and
    \inference{
    p \in \hist
    }{
    \lkw{if}\; p\; \lkw{then}\; s_1\; \lkw{else}\; s_2 \vbar \cfg \pstep s_1 \vbar \cfg
    }[(\text{if-t})]
    \and
    \inference{
    p \notin \hist
    }{
    \lkw{if}\; p\; \lkw{then}\; s_1\; \lkw{else}\; s_2 \vbar \cfg \pstep s_2 \vbar \cfg
    }[(\text{if-f})]
    \and
    \inference{
    \repos(r) = \{(w_1, p_1), \ldots, (w_k, p_k)\} \\
    s_i \triangleq s[w_i, p_i / w, p] \quad k \geq 1
    }{
    \lkw{for}\; (w, p)\; \lkw{in}\; r\; \{\, s \,\} \vbar (\env, \gst) \pstep
        s_1 \,;\, \cdots \,;\, s_k \vbar (\env, \gst)
    }[(\text{for})]
    \and
    \inference{
    \repos(r) = \emptyset
    }{
    \lkw{for}\; (w, p)\; \lkw{in}\; r\; \{\, s \,\} \vbar (\env, \gst) \pstep \lkw{skip} \vbar (\env, \gst)
    }[(\text{for-emp})]
    \end{mathpar}

    \boxed{
    \mathcal{T}, \gst \rstep \mathcal{T'}, \gst'
    }
    \begin{mathpar}
    \inference{
    s \vbar (\env, (\hist, \wbuf, \repos)) \pstep s' \vbar (\env', (\hist', \wbuf', \repos'))
    }{
    \mathcal{T} \cup \{(s, \env)\}, (\hist, \wbuf, \repos) \rstep \mathcal{T} \cup \{(s', \env')\}, (\hist', \wbuf', \repos')
    }[(\text{exec})]
    \and
    \inference{
    \textit{init}_\pi(s, \env)
    }{
    \mathcal{T}, (\hist, \wbuf, \repos) \rstep \mathcal{T} \cup \{(s, \env)\}, (\hist, \wbuf, \repos)
    }[(\text{user-new-await})]
    \and
    \inference{
    p \in \pext \quad p \notin \hist
    }{
    \mathcal{T}, (\hist, \wbuf, \repos) \rstep \mathcal{T}, (\hist \cup \{p\},\; \wbuf,\; \repos)
    }[(\text{env-trigger})]
    \and
    \inference{
    w \in \wbuf \quad \tau(w) = (s, \env)
    }{
    \mathcal{T}, (\hist, \wbuf, \repos) \rstep \mathcal{T} \cup \{(s, \env)\},\; (\hist, \wbuf \setminus \{w\}, \repos)
    }[(\text{sched})]
    \end{mathpar}
    \caption{Operational semantics for $\lang$: statement small-step rules (top) and runtime configuration rules (bottom). Standard rules such as sequencing or $\lkw{skip}$ are omitted; the full definition is in Appendix~\ref{app:opsem-full}.}
    \label{fig:lang-prog-step}
\end{figure}

Figure~\ref{fig:lang-prog-step} defines the operational semantics for $\lang$.
$\lkw{trigger}$, $\lkw{insert}$, $\lkw{take}$, and $\lkw{wake}$ are atomic effectful statements that update the global state and reduce to $\lkw{skip}$.
$\lkw{if}$ statements branch on whether a predicate is in the monotonic history $\Sigma$\footnote{$\lang$ only allows branching on single monotonic predicates; extending this to combinations as well as more-standard program expressions would be straightforward.}.
$\lkw{for}$ statements iterate over every $(w, p)$ in $\repos(r)$ present \emph{at the start of the for-loop}.
We omit other conventional rules for $\lkw{loop}$, $\lkw{skip}$, etc.
The full definition is in \appref{opsem-full}.

As explained in \secref{overview}, a Rust async runtime in general has many pollables and reactors currently executing, and additionally interacts with other trusted components, user tasks and the external environment, none of which are directly modeled by $\lang$ programs.
We model them via additional non-deterministic transitions interleaved with the execution of statements defined in \figref{lang-prog-step}.

We define \emph{runtime configurations} as a pair $(\mathcal{T}, (\hist, \wbuf, \repos))$ where $\mathcal{T}$ is a multiset of pairs $\Stmt \times \Env$ representing all currently running pollables and reactors. The runtime-level judgment $\mathcal{T}, (\hist, \wbuf, \repos) \rstep \mathcal{T'}, (\hist', \wbuf', \repos')$ is the reflexive transitive closure of the rules at the bottom of \figref{lang-prog-step}.
Rule (exec) models normal execution of a single-step of a currently-running pollable (or reactor). Rule (user-new-await) models a new pollable instance being started due to the compilation of an \rcode{await} usage in a user-task. We abstract over the details of initializing a new instance of some pollable $\pi$ (\eg{} initializing the environment) using the $\textit{init}_\pi$ predicate on the new pair of its implementation $s$ and environment $\env$. For simplicity, we do not model dynamic creation of reactors but assume these to be present in an initial runtime configuration.

Rule (env-trigger) adds an external predicate to the history nondeterministically; this models that the environment may make an external condition true at arbitrary time. Rule (sched) schedules a waker $w$ in the rescheduling queue $\wbuf$ by reinvoking the corresponding pollable.
Since our model abstracts over the user task code, which (due to the Rust compiler's transformations) is guaranteed to re-call the awaited runtime pollable, our $\tau$ function formally maps the waker \emph{directly} to this runtime pollable's representation, which is made available for rescheduling by our rules. In the real runtime system, it is the user task that originally awaited the pollable that would be rescheduled by the waker.

\textbf{Obligations.}
A key intuition for what is needed for an async runtime to be responsive is that the runtime \emph{never forgets} async tasks that may be able to make further progress (for example, it doesn't forget a task waiting on a timer).
To formalize this idea, we introduce the concept of \emph{obligations}: responsibilities that runtime components (especially reactors) are conceptually assigned that must not be forgotten until they are discharged.

Specifically, we use pairs $(w, p) \in \Waker \times \pred$ in our novel operational semantics to represent the obligation: \emph{if the monotonic condition $p$ ever holds, the runtime must eventually resume the task associated with $w$}.
An obligation is created every time a user task awaits on a pollable, represented by the waker used for resumption of the task and the condition being waited on. If the pollable waited on returns $\lReady$, the obligation is discharged as the user code will be returned to directly; on the other hand, if a reactor executes $\lkw{wake}(w)$ this discharges the obligation based on our assumptions on the fair scheduler. These are the only ways these obligations can be discharged.

Given a repository map, $\repos$, the \emph{obligations of a reactor} $\reactor$, written $\Ob{\reactor}$ are defined as the union of the obligations in $\repos(r)$ for every repository $r$ owned by $\reactor$.
For a given global state $\gst$ we overload this syntax to define the \emph{total obligations} of a global state $\mathit{Ob}(\gst)$ to be the union of $\Ob{\reactor}$ over all reactors $\reactor$.
Note that $\Ob{\reactor}$ grows when a pollable transfers a waker via $\lkw{insert}$ to $\reactor$ and shrinks when one is removed via $\lkw{take}$.

\subsection{Eventual Progression and Component Local Conditions}
\label{sec:model-properties}

We now state an overall guarantee in terms of the interaction between user code and an async runtime; we use this high-level notion to derive decomposed verification conditions specific to pollable and reactor implementations.

\begin{definition}[Eventual Progression]
\label{def:eventual-progression}
An async runtime satisfies Eventual Progression iff, for any runtime configuration $(\mathcal{T}, \gst)$ reached during its execution, and for all user tasks awaiting on a pollable $s$ with predicate $p$, either (a) $s$ is running with $(s,\env) \in \mathcal{T}$ or (b) for which a waker $w$ exists with $(w,p)\in \Ob{\gst}$, it is guaranteed that if, on further execution $p$ eventually becomes true, then eventually the user task will be resumed.
\end{definition}
Case (a) corresponds to a user task awaiting a pollable which is already running; this can potentially either result in delegation to a reactor (and case (b)) or returning a $\lReady(v)$ value directly causing the user task to resume. This allows us to define conditions specific to the implementation of a single pollable:
\begin{definition}[Pollable Responsiveness]
\label{def:pollable-responsiveness}
A pollable $\pi$ is \emph{responsive} if every invocation with obligation $(w, p)$ satisfies:
\begin{enumerate}
    \item[(P1)] if the invocation returns $\lReady(v)$, then $p \in \hist$ at the return point;
    \item[(P2)] if $p \in \hist$ at the entry of the invocation, then the invocation returns $\lReady(v)$ for some $v$;
    \item[(P3)] if the invocation returns $\lPending$, then either $(w, p) \in \repos(r)$ for some \public{} repository $r$, or $w \in \wbuf$ at the return point;
\end{enumerate}
\end{definition}
Definition~\ref{def:pollable-responsiveness} captures two key concerns.
(P1) and (P2) form the soundness and completeness for completion: (P1) ensures $\lReady$-returns reflect actual completion; (P2) ensures that an already-completed computation is eventually reported.
(P3) ensures obligation-preservation: when a pollable yields, the obligation it carries must not be forgotten by the runtime. It may be transferred to a \public{} repository (via $\lkw{insert}$) or discharged immediately by waking $w$ (via $\lkw{wake}$).

Considering Case (b) in \defref{eventual-progression}, we must require enough of a reactor to also ensure that obligations are preserved and eventually discharged when ready. We express these conditions per iteration $i$ of the $\lkw{loop}\;s$ construct defining the reactor, and write $\gst_{\mathit{start}}^i, \gst_{\mathit{end}}^i$ for the global states at iteration start and end.
In the following, $\mathit{Ob}^i_{\mathit{con}}$ denotes the obligations inserted into \public{} repositories \emph{during} iteration $i$ (\eg{} by concurrently-running pollables). We write $\mathit{Wake}_i \triangleq \{\, w \mid \lkw{wake}(w) \text{ fires during iteration } i \,\}$ for the set of all obligations discharged via $\lkw{wake}$ during the iteration:

\begin{definition}[Reactor Responsiveness]
\label{def:reactor-responsiveness}
A reactor $\lkw{loop}\;s$ is \emph{responsive} if every iteration $i$ satisfies:
\begin{align*}
    \text{(R1)} \quad & \forall (w, p) \in \mathit{Ob}(\gst_{\mathit{start}}^i).\; p \in \hist_{\mathit{start}}^i \;\Rightarrow\; w \in \mathit{Wake}_i \\
    \text{(R2)} \quad & \mathit{Ob}(\gst_{\mathit{end}}^i) \;\supseteq\; \bigl(\mathit{Ob}(\gst_{\mathit{start}}^i) \cup \mathit{Ob}^i_{\mathit{con}}\bigr) \setminus \bigl\{(w, p) \mid w \in \mathit{Wake}_i\bigr\}
\end{align*}
\end{definition}
(R1) guarantees progress at each iteration; the obligations whose predicate already holds must be woken in the \emph{same} iteration. In principle this condition could be weakened, but in practice this strong guarantee is common.
(R2) guarantees that no obligation is lost; obligations not discharged during the iteration must be retained in the \public{} repositories by the end of the iteration.
Obligations arriving mid-iteration via $\mathit{Ob}^i_{\mathit{con}}$ are not required to be woken in the same iteration even if their predicate already holds by iteration end; (R2) carries them into the next iteration.

The relationship between these component local conditions and the global Eventual Progression property is summarized as follows; the corresponding soundness theorem and its proof are in Section~\ref{sec:verification-soundness}.

\begin{theorem}[Componentwise Decomposition]
\label{thm:decomposition-informal}
Assume the three trusted components of Section~\ref{sec:overview-ep}. If every pollable in the runtime preserves obligations when yielding (Definition~\ref{def:pollable-responsiveness}, (P3)) and every reactor is responsive (Definition~\ref{def:reactor-responsiveness}), then the runtime satisfies Eventual Progression (Definition~\ref{def:eventual-progression}).
\end{theorem}

The focus of Section~\ref{sec:verification} is to define new and efficient \emph{modular} static analyses whose success implies these required properties. We design abstract domains and tailored verification conditions $\VCp$ (for pollables) and $\VCR$ (for reactors) that, when discharged on the syntactic body of a component, imply that the component is responsive.

 \section{Static Analysis of Eventual Progression}
\label{sec:verification}

This section defines novel static analyses presented in an abstract-interpretation style that verify the componentwise properties defined and motivated in the previous section.
Our abstract domains are novel in that they focus on the notions of monotonic predicates and undischarged obligations.
Section~\ref{sec:verification-domain} introduces a novel unified abstract domain shared by both analyses; Sections~\ref{sec:verification-pollable} and~\ref{sec:verification-reactor} define analysis-refined pollable and reactor verification conditions; Section~\ref{sec:verification-soundness} states the soundness theorem and lifts the componentwise guarantees to a user-facing runtime-responsiveness theorem.

\subsection{Abstract Domain}
\label{sec:verification-domain}
We design an abstract domain which tracks \emph{two} kinds of information at each program point: observations on monotonic predicates, and \emph{symbolic obligations} abstracting over sets of concrete obligations held by reactors (their repositories).

We define abstract states $\sigma = (S, \srepos, \sloc) \in \AbsState$, where $S$ is a map tracking \emph{predicate observations}, $\srepos$ is a map symbolically tracking obligations stored in \public{} repositories, and $\sloc$ a set tracking locally-held symbolic obligations.
We explain each component formally below.

States are ordered pointwise.
The join $\sqcup$ takes the pointwise lub of $S$, pointwise union of $\srepos$, and union of $\sloc$.
The bottom element $\bot_{\AbsState}$ represents an \emph{unreachable} program state.
$\bot_{\AbsState} \sqcup \sigma = \sigma$ for any abstract state.

\textbf{Predicate observations.}
We use the set $\Obs = \{\bot, \atrue, \afalse, \top\}$ for \emph{predicate observations}. $\atrue$ means $p$ \emph{has been observed} to hold; $\afalse$ means $p$ was observed not to hold at an earlier moment; $\top$ is unknown; $\bot$ means unreachable.

Two operators are defined over $\Obs$: the refinement operator $\refop$ updates an observation along a single execution path (defined below); the join $\sqcup$ merges observations from multiple paths, taken as the least upper bound on the ordering $\bot \sqsubseteq \atrue, \afalse \sqsubseteq \top$ (where $\atrue$ and $\afalse$ are incomparable and $\atrue \sqcup \afalse = \top$).

The monotonic refinement operator $\refop : \Obs \times \{\atrue, \afalse\} \to \Obs$ updates a prior observation by a fresh observation result (e.g., a branch of an if-statement) along a single path, encoding the monotonicity of predicates:
\begin{equation*}\label{eq:directional-refine}
\begin{aligned}
\atrue \refop \atrue &= \atrue
    & \qquad \afalse \refop \atrue &= \atrue
    & \qquad \top \refop v &= v \\
\atrue \refop \afalse &= \bot
    & \qquad \afalse \refop \afalse &= \afalse
    & \qquad \bot \refop v &= \bot \\
\end{aligned}
\end{equation*}
If the value of a predicate is unknown $\top$, it can be refined into a concrete observation (either $\atrue$ or $\afalse$).
A $\afalse$ observation can be refined to $\atrue$, while refining $\atrue$ into $\afalse$ violates monotonicity, indicating the current path is unreachable.

$\bot$ cannot be refined to any other values.
We adopt a smash-product convention for abstract states: if $S(p) = \bot$ for any $p$, the entire state $\sigma$ is identified as the bottom $\bot_{\AbsState}$.
Intuitively, this means any observation that violates monotonicity reveals an infeasible path; our analyses rely on this insight to prune infeasible paths.

\textbf{Symbolic obligations.}
Let $\sWaker$ be a set of \emph{symbolic wakers}.
A \emph{symbolic obligation} is a triple $\hat{o} = (\hat{w}, \hat{p}, \tau) \in \sOb \triangleq \sWaker \times \pred \times \{\atrue, \top\}$.
We use tag $\tau$ to distinguish different assumptions about predicate $\hat{p}$'s state at the start of analysis: tag $\atrue$ indicates $\hat{p}$ is assumed to hold; $\top$ indicates $\hat{p}$'s state is unknown.
$\srepos : \Repo \to \wp(\sOb)$ maps each \public{} repository to the symbolic obligations currently stored in it; $\sloc \in \wp(\sOb)$ collects the symbolic obligations held locally (taken out of a repository, or initialized at the start of the analysis).

\begin{figure}[t]
    \raggedright
    \boxed{
    \tfn[s] : \AbsState \to \AbsState
    }
    \begin{align*}
    \tfn[\lkw{skip}](\sigma) &= \sigma \\
    \tfn[\lkw{trigger}(p)](\sigma) &= \sigma\bigl[\,S \mathrel{:=} S[p \mapsto S(p) \refop \atrue]\,\bigr] \\
    \tfn[\lkw{insert}(r, \hat{w}, \hat{p})](\sigma)
        &= \sigma\bigl[\,\srepos(r) \mathrel{:=} \srepos(r) \cup \{(\hat{w}, \hat{p}, \tau)\},\;\; \sloc \mathrel{:=} \sloc \setminus \{(\hat{w}, \_, \_)\}\,\bigr]
        \\[-2pt] &\qquad \text{if } (\hat{w}, \hat{p}, \tau) \in \sloc
        \;\land\; \hat{p} \in \mathit{img}(\cp_{r}) ;
        \text{ otherwise } \bot_{\AbsState} \\
    \tfn[\lkw{wake}(\hat{w})](\sigma)
        &= \sigma\bigl[\,\sloc \mathrel{:=} \sloc \setminus \{(\hat{w}, \_, \_)\}\,\bigr]
        \\[-2pt] &\qquad \text{if } (\hat{w}, \_, \_) \in \sloc ; \text{ otherwise } \bot_{\AbsState} \\
    \tfn[\lkw{if}\; p\; \lkw{then}\; s_1\; \lkw{else}\; s_2](\sigma)
        &= \sigma_{s_1} \sqcup \sigma_{s_2}\\
        \text{where } \sigma_{s_1} & = \tfn[s_1]\bigl(\sigma\bigl[\,S \mathrel{:=} S[p \mapsto S(p) \refop \atrue]\,\bigr]\bigr) \\
        \sigma_{s_2} & = \tfn[s_2]\bigl(\sigma\bigl[\,S \mathrel{:=} S[p \mapsto S(p) \refop \afalse]\,\bigr]\bigr) \\
    \tfn[s_1 \,;\, s_2](\sigma) &= \tfn[s_2](\tfn[s_1](\sigma)) \\
    \tfn[\lkw{return}\; v](\sigma) &= \bot_{\AbsState}
    \end{align*}
    \caption{
        Transfer functions for the shared statements of $\lang$.
        All transfer functions preserve $\bot_{\AbsState}$, i.e., $\tfn[s](\bot_{\AbsState}) = \bot_{\AbsState}$ for every $s$.
}
    \label{fig:lang-transfer}
\end{figure}

\begin{figure}[t]
\begin{align*}
    \tfn[\lkw{take}(r, \hat{w})](\sigma)
        &= \sigma\bigl[\,\srepos(r) \mathrel{:=} \srepos(r) \setminus \{(\hat{w}, \hat{p}, \tau)\},\;\; \sloc \mathrel{:=} \sloc \cup \{(\hat{w}, \hat{p}, \tau)\}\,\bigr]
        \\[-2pt] &\qquad \text{if } (\hat{w}, \hat{p}, \tau) \in \srepos(r); \text{ otherwise } \bot_{\AbsState} \\
    \tfn[\lkw{for}\;(w, p)\;\lkw{in}\; r\; \{\, s \,\}](\sigma)
        &= \tfn[s_k](\,\cdots\, \tfn[s_1](\sigma))
        \\[-2pt] &\qquad \text{where } \srepos(r) = \{(\hat{w}_1, \hat{p}_1, \tau_1), \ldots, (\hat{w}_k, \hat{p}_k, \tau_k)\}
        \\[-2pt] &\qquad \text{and } s_i \triangleq s[\hat{w}_i, \hat{p}_i / w, p].
    \end{align*}
    \caption{Transfer functions for reactor-only $\lang$ statements.}
    \label{fig:reactor-transfer}
\end{figure}

\textbf{Transfer functions.}
Figure~\ref{fig:lang-transfer} defines the transfer function $\tfn[s]$ for $\lang$ statements that are shared between pollables and reactors.
$\lkw{trigger}(p)$ refines $S(p)$ to $\atrue$.
$\lkw{insert}$ moves an obligation from $\sloc$ into $\srepos(r)$, requiring the obligation to be held locally and matching the repository's typing.
$\lkw{wake}$ consumes a locally-held obligation.
$\lkw{if}\; p$ refines $S(p)$ on each branch via $\refop$ and joins the two results after the statement.
$\lkw{return}\;v$ yields $\bot_{\AbsState}$, as the execution is terminated.

$\lkw{insert}$, $\lkw{take}$, and $\lkw{wake}$ yield $\bot_{\AbsState}$ when side-conditions fail; this models infeasible paths due to \eg{} type mismatches or exceptions.

Figure~\ref{fig:reactor-transfer} extends the transfer functions to reactor-only statements $\lkw{take}$ and $\lkw{for}$.
$\lkw{take}$ moves an obligation out of a \public{} repository and holds it locally;
$\lkw{for}\;(w, p)\;\lkw{in}\;r$ is unrolled for every symbolic obligation in $\srepos(r)$.
The $w$ and $p$ mentioned in the loop body are substituted with the $\hat{w}_i$ and $\hat{p}_i$ of each obligation.

\subsection{Pollable Analysis Verification Conditions}
\label{sec:verification-pollable}

We analyze each pollable implementation $\pi$ \emph{twice}, running our transfer function with two different initial states, defined as follows (in which $\hat{w}$ and $\hat{p}$ are symbolic waker and completion predicates):
\begin{itemize}
    \item $\sigma^{\top}_{\mathit{start}}$: $S_{\mathit{start}}(\hat{p}) = \top$, $\sloc_{\mathit{start}} = \{(\hat{w}, \hat{p}, \top)\}$, $\srepos_{\mathit{start}}(r) = \emptyset$ for all $r$;
    \item $\sigma^{\atrue}_{\mathit{start}}$: $S_{\mathit{start}}(\hat{p}) = \atrue$, $\sloc_{\mathit{start}} = \{(\hat{w}, \hat{p}, \atrue)\}$, otherwise identical.
\end{itemize}
For a well-formed pollable body, every path terminates with a $\lkw{return}$, so the abstract state immediately before each $\lkw{return}$ summarizes what is known about $\hist$ and the entry obligation at that exit point.

\textbf{Verification conditions.}
$\VCp(\pi)$ checks the abstract state immediately before each return site of $\pi$:
\begin{itemize}
    \item \text{(P1$_{vc}$):} under $\sigma^{\top}_{\mathit{start}}$, at every $\lkw{return}\;\lReady(v)$: $S(\hat{p}) = \atrue$;
    \item \text{(P2$_{vc}$):} under $\sigma^{\atrue}_{\mathit{start}}$, every reachable return site is a $\lReady$-return;
    \item \text{(P3$_{vc}$):} under $\sigma^{\top}_{\mathit{start}}$, at every $\lkw{return}\;\lPending$: $\sloc_{\mathit{end}} = \emptyset$.
\end{itemize}
The three checks discharge (P1), (P2), and (P3) of Definition~\ref{def:pollable-responsiveness}, respectively.
($\text{P3}_{vc}$) witnesses that the locally-held obligation has been transferred to a \public{} repository (via $\lkw{insert}$) or discharged (via $\lkw{wake}$) by the time the pollable yields.

\paragraph{Example:}
The Timer pollable in Section~\ref{sec:overview-timer} can be modeled in $\lang$ as
\begin{align*}
    \pi_{\mathit{timer}} \;\triangleq\; &
    \lambda(w, p_c).\; \\
    & \lkw{insert}(\mathtt{A}, w, p_c) \,;\; \\
    & \lkw{if}\; p_c\; \lkw{then} \;\{\; \lkw{return}\;\lReady(()) \;\}\; \lkw{else} \;\{\; \lkw{return}\;\lPending \;\}\;
\end{align*}
where $p_c \in \pext$ is the timer's completion predicate (i.e., \rcode{self.instant <= now()}) and $\mathtt{A}$ is the Reactor's \public{} repository.

When running the analysis with $\sigma^{\top}_{\mathit{start}}$,
$\lkw{insert}(\mathtt{A}, w, p_c)$ moves the entry obligation $(\hat{w}, p_c, \top)$ from $\sloc$ to $\srepos(\mathtt{A})$, leaving $\sloc = \emptyset$.
The $\lkw{if}$ statement refines the state of $p_c$ to $S(p_c) = \atrue$ on the then-branch and $S(p_c) = \afalse$ on the else-branch.
On the then-branch we reach $\lkw{return}\;\lReady(())$ with $S(p_c) = \atrue$, discharging ($\textsc{P1}_{\mathrm{vc}}$).
On the else-branch we reach $\lkw{return}\;\lPending$ with $\sloc_{\mathit{end}} = \emptyset$, discharging ($\textsc{P3}_{\mathrm{vc}}$).

When running the analysis with $\sigma^{\atrue}_{\mathit{start}}$: the else-branch refinement gives $S(p_c) = \atrue \refop \afalse = \bot$, so only the $\lReady$-return is reachable, discharging (P2).
$\VCp(\pi_{\mathit{timer}})$ holds.
\subsection{Reactor Analysis Verification Conditions}
\label{sec:verification-reactor}

For each \public{} repository $r$ of type $(T_r, \cp_r)$, we analyze the reactor implementation from an initial $\srepos_{\mathit{start}}(r)$ with \emph{two} symbolic obligations:
\begin{displaymath}
    (\hat{w}^{\atrue}_r,\, \hat{p}^{\atrue}_r,\, \atrue),
    \qquad
    (\hat{w}^{\top}_r,\, \hat{p}^{\top}_r,\, \top),
\end{displaymath}
where $\hat{p}^{\atrue}_r, \hat{p}^{\top}_r \in \mathit{img}(\cp_r)$.
The initial observation function is $S_{\mathit{start}}(\hat{p}^{\atrue}_r) = \atrue$ and $S_{\mathit{start}}(\hat{p}^{\top}_r) = \top$; $\sloc_{\mathit{start}} = \emptyset$.
The $\atrue$-tagged obligation models the case where the completion predicate already holds before the iteration (for checking (R1)); the $\top$-tagged obligation models the case where the predicate's state is unknown (for checking (R2)).
Given a reactor body $\lkw{loop}\;s$, the analysis runs \emph{once} on $s$ with the initial state $\sigma_{\mathit{start}}$.
Tt computes the end-of-iteration state $\sigma_{\mathit{end}} = \tfn[s](\sigma_{\mathit{start}})$ and no fixpoint over $\lkw{loop}$ is needed.

\textbf{Verification conditions.}
$\VCR(s)$ inspects $\sigma_{\mathit{end}}$ and quantifies over the symbolic obligations seeded at iteration start.
\begin{itemize}
    \item ($\text{R1}_{vc}$): for every $\hat{o} = (\hat{w}, \hat{p}, \atrue) \in \srepos_{\mathit{start}}(r)$, $\hat{o} \notin \sloc_{\mathit{end}}$ and $\hat{o} \notin \bigcup_{r'} \srepos_{\mathit{end}}(r')$.
    \item ($\text{R2}_{vc}$): for every $\hat{o} = (\hat{w}, \hat{p}, \top) \in \srepos_{\mathit{start}}(r)$, $\hat{o} \notin \sloc_{\mathit{end}}$.
\end{itemize}
$\VCR(s) \triangleq \text{R1}_{vc} \wedge \text{R2}_{vc}$.
$\text{R1}_{vc}$ witnesses that all obligations whose predicates hold at iteration start have been consumed by $\lkw{wake}$ and removed from \public{} repositories.
$\text{R2}_{vc}$ witnesses that obligations with unknown predicates are either discharged by $\lkw{wake}$ (hence absent from $\sloc$ and every $\srepos$) or preserved in some \public{} repository for future iterations.

\paragraph{Example:}
The Timer Reactor of \secref{overview-timer} can be encoded in $\lang$ as
\begin{align*}
    \reactor_{\mathit{timer}}\; \triangleq\;
    \lkw{reactor}\;\{\;\;
        & \lkw{shared}\;\mathtt{A} : (\mathit{Timer}, \cp_{\mathtt{A}})\;\lkw{in} \\
        & \lkw{loop}\;\;\lkw{for}\;(w, p)\;\lkw{in}\;\mathtt{A}\;\{ \\
        & \quad \lkw{if}\;p\;\lkw{then}\;\{\,\lkw{take}(\mathtt{A}, w)\,;\,\lkw{wake}(w)\,\}\;\lkw{else}\;\lkw{skip} \,\}\}
\end{align*}

We initialize the repository map $\srepos_{\mathit{start}}$ with the two symbolic obligations introduced above, i.e., $\srepos_{\mathit{start}}(\mathtt{A}) = \{\hat{o}^{\atrue},  \hat{o}^{\top}\}$ and set the local obligation set to empty $\sloc_{\mathit{start}} = \emptyset$.

The $\lkw{for}$-rule instantiates the body once per symbolic obligation.
For the instantiation with $\hat{o}^{\atrue}$, since we assumed $S(\hat{p}^{\atrue}) = \atrue$ at iteration start, the else-branch refines to $\bot_{\AbsState}$; only the then-branch survives.
$\lkw{take}(\mathtt{A}, \hat{w}^{\atrue})$ moves $\hat{o}^{\atrue}$ into $\sloc$ and subsequently $\hat{o}^{\atrue}$ is consumed by $\lkw{wake}(\hat{w}^{\atrue})$.
After the if-statement, $\srepos(\mathtt{A}) = \{\hat{o}^{\top}\}$ and $\sloc = \emptyset$.

For the instantiation with $\hat{o}^{\top}$, both branches survive; the join unions $\srepos$ and $\sloc$.
After the for-loop, we have $\srepos_{\mathit{end}}(\mathtt{A}) = \{\hat{o}^{\top}\}$ (contributed by the else-branch) and $\sloc_{\mathit{end}} = \emptyset$.

We see that the symbolic obligation $\hat{o}^{\atrue}$ is not in $\sloc_{\mathit{end}}$ or $\srepos_{\mathit{end}}(\mathtt{A})$ by the end of the iteration, hence $\text{R1}_{vc}$ holds.
Meanwhile, obligation $\hat{o}^{\top}$ only exists in $\srepos_{\mathit{end}}(\mathtt{A})$, hence $\text{R2}_{vc}$ holds and the reactor is verified.

\paragraph{Example (buggy):}
Now consider a buggy implementation of the same reactor obtained by replacing the then-branch's $\lkw{wake}$ with $\lkw{skip}$ (the reactor takes the obligation out of $\mathtt{A}$ but forgets to wake it).
The iteration with obligation $\hat{o}^{\atrue}$ never discharges the obligation in the then-branch, leaving $\hat{o}^{\atrue} \in \sloc_{\mathit{end}}$, hence failing $\text{R1}_{vc}$.

\subsection{Soundness}
\label{sec:verification-soundness}

In this subsection, we state the key soundness theorems and provide proof sketches for them.
We first show that the verification conditions soundly approximate the responsiveness properties for pollables and reactors as we defined in Definition~\ref{def:pollable-responsiveness} and Definition~\ref{def:reactor-responsiveness}.

\begin{theorem}[Pollable Soundness]
\label{thm:pollable-soundness}
Let $\pi$ be a pollable in $\lang$. If $\VCp(\pi)$ holds, then $\pi$ is responsive (Definition~\ref{def:pollable-responsiveness}).
\end{theorem}

\paragraph{Proof sketch.}
The transfer functions defined in Fig.~\ref{fig:lang-transfer} \emph{simulates} the concrete semantics in Fig.~\ref{fig:lang-prog-step}: the abstract state at every program point over-approximates the concrete state.

(P1) follows because $\VCp$ enforces $S(\hat{p}) = \atrue$ at every $\lReady$-return site, which by simulation gives $p \in \hist$ in every concrete trace.
(P2) follows because, when $S(\hat{p})$ is initialized to $\atrue$, $\VCp$ rules out reaching any $\lPending$-return site, so concrete executions starting with $p \in \hist$ at entry must return $\lReady$.
(P3) follows because $\VCp$ enforces $\sloc_{\mathit{end}} = \emptyset$ at every $\lPending$-return site; an obligation can only leave $\sloc$ either via $\lkw{insert}$ (placing it in $\srepos$) or $\lkw{wake}$ (firing $w$), so by simulation the concrete trace either transfers $(w, p)$ to a \public{} repository or fires $w$ before returning $\lPending$.

\begin{theorem}[Reactor Soundness]
\label{thm:reactor-soundness}
Let $\reactor$ be a reactor in $\lang$ whose body is $\lkw{loop}\;s$. If $\VCR(s)$ holds, then $\reactor$ is responsive (Definition~\ref{def:reactor-responsiveness}).
\end{theorem}

\paragraph{Proof sketch.}
By simulation, $\sigma_{\mathit{end}} = \tfn[s](\sigma_{\mathit{start}})$ over-approximates every concrete iteration-end configuration.
The symbolic obligation $\hat{o}^{\atrue}$ over-approximates all obligations whose monotonic predicates hold at the start of an iteration.
$\text{R1}_{vc}$ implies that the iteration both removes all corresponding obligations from every \public{} repository (via $\lkw{take}$) and consumes them locally (via $\lkw{wake}$), so the concrete wakers are in $\mathit{Wake}_i$ (the set of wakers being woken up in iteration $i$); therefore (R1) holds.

$\hat{o}^{\top}$ over-approximates both obligations present at iteration start (preserved from previous iterations) and those inserted concurrently during the iteration.
$\text{R2}_{vc}$ implies that, on every path, such concrete obligations are either discharged via $\lkw{wake}$ or preserved in some \public{} repository, satisfying (R2).

\begin{theorem}[Reactor Eventual Progression]
\label{thm:reactor-eventual-progression}
Let $\lkw{loop}\;s$ be a verified reactor and $(w, p)$ an obligation ever placed in one of its \public{} repositories; the following holds
\[ \square\bigl(\, p \in \hist \;\Rightarrow\; \Diamond\, w \in \wbuf \,\bigr). \]
\end{theorem}

\paragraph{Proof sketch.}
Theorem~\ref{thm:reactor-soundness} gives that verified reactors satisfy Definition~\ref{def:reactor-responsiveness}.
By (R2) of Def.~\ref{def:reactor-responsiveness}, $(w, p)$ is preserved at every iteration end, so it will remain in some \public{} repository from when it was first inserted until $\lkw{wake}(w)$ fires.
By monotonicity of the predicate $p$, once $p \in \hist$ it remains so; since the reactor iterates infinitely often, there will be an iteration $j$ at whose start $p \in \hist$, and (unless $\lkw{wake}(w)$ has already been called) $(w, p)$ will be still present in some repository.
By (R1) of Def.~\ref{def:reactor-responsiveness}, $w$ will be woken up at iteration $j$, i.e., $w \in \mathit{Wake}_j$, and by the (wake) rule in Fig.~\ref{fig:lang-prog-step}, eventually $w \in \wbuf$.
This discharges Definition~\ref{def:eventual-progression} case (b).

Theorems~\ref{thm:pollable-soundness} and~\ref{thm:reactor-eventual-progression} together establish the runtime side of Eventual Progression: every obligation introduced into the runtime, whose monotonic predicate eventually holds, eventually has its waker fired.
Combined with the three trusted components of Section~\ref{sec:overview-ep} (a fair scheduler, a correct compiler, and a responsive environment), the runtime delivers the end-to-end responsiveness property below.

\begin{theorem}[Runtime Responsiveness]
\label{thm:runtime-responsiveness}
Suppose every reactor and every pollable in the runtime satisfies its verification condition, async-task compilation is faithful, the scheduler is fair, and the external predicates a task depends on eventually hold. Then every async task running in this runtime eventually makes progress.
\end{theorem}

\paragraph{Proof sketch.}
Faithful compilation transforms every async function into a state-machine pollable whose \rcode{poll} method delegates to the awaitees' \rcode{poll} method.

By induction on the depth of nested $\lkw{await}$s, the base case (a task awaiting a runtime pollable) follows from Theorems~\ref{thm:pollable-soundness} and~\ref{thm:reactor-eventual-progression}.

For the inductive case (a task $t$ awaiting a sub-task $t'$), $\pi_t$'s \rcode{poll} invokes $\pi_{t'}$'s \rcode{poll}.
The inductive hypothesis gives that $\pi_{t'}$ eventually returns $\lReady$; faithful compilation propagates the result to $\pi_t$, which continues past the await.

\begin{corollary}[Responsiveness Attribution]
\label{cor:responsiveness-attribution}
If an async task is observed to hang at a program point, at least one of the trusted components must have failed, or the user's code itself has unintended non-termination or deadlocks independently of the runtime's pollables. The verified pollable and reactor implementations provided by the async runtime \emph{cannot} be the cause.
\end{corollary}
The corollary above is the contrapositive of Theorem~\ref{thm:runtime-responsiveness}.
In practice, our analyses rule out async runtime's implementations as the cause of responsiveness issues.

\subsection{Extension 1: Terminating Reactors and Precedes Relations}
\label{sec:terminating-reactors}

Our model so far requires a reactor to run an infinite loop over its repositories, which is the most-common idiom. In this subsection we show an adaptation of our techniques to support reactor-like components which handle only \emph{finitely many} wakers, and can thus justify termination.

This idiom of \emph{terminating reactors} arises in practice \eg{} in the implementation of the receiving end of a one-shot channel.
\begin{figure}[tbp]
    \centering
    \begin{align*}
    \text{Reactors}   \quad & \reactor ::= \ldots
                              \;\mid\; \lkw{t-reactor}\;\{\,
                                \lkw{shared}\;\vv{r_s : (T_{r_s}, \cp_{r_s})}\;
                                \lkw{in}\; s
                              \,\} \quad (\mathit{wfr}(s))
    \end{align*}
    \begin{align*}
    \tfn[\lkw{insert}(r, \hat{w}, \hat{p})](\sigma) &= \sigma\bigl[\,
        S(\mathit{tfr}) \mathrel{:=} \atrue,\;\;
        \srepos(r) \mathrel{:=} \srepos(r) \cup \{(\hat{w}, \hat{p}, \tau)\}, \\[-2pt]
        &\qquad\phantom{\sigma\bigl[\,}
        \sloc \mathrel{:=} \sloc \setminus \{(\hat{w}, \_, \_)\}
        \,\bigr] \\[-2pt]
        &\qquad \text{if } (\hat{w}, \hat{p}, \tau) \in \sloc
        \;\land\; \hat{p} \in \mathit{img}(\cp_{r}) ;
        \text{ otherwise } \bot_{\AbsState} \\[3pt]
    \sigma|_{p \mapsto \afalse} &= \sigma\bigl[\,
        S(p) := S(p) \refop \afalse,\;\;
        {\precedes} := (\precedes \,\cup\, \{(q,p) : q \neq p,\; S(q) = \atrue\})^{+}
        \,\bigr]
    \end{align*}
    \caption{Extensions for terminating reactors. A new $\lkw{t-reactor}$ syntax construct. Two new cases of the transfer function for the strict-obligation-transfer (SOT) extension: $\lkw{insert}$ records the transfer event $\mathit{tfr}$, and refining a predicate to $\afalse$ extends $\precedes$ ($X^{+}$ denotes transitive closure). The state smashes to $\bot$ if $S(p) \refop \afalse = \bot$; all other forms are unchanged.}
    \label{fig:lang-syntax-ext}
    \label{fig:strict-transfer}
\end{figure}
The receiver is an async future awaiting a value, while its peer code \rcode{send} is an ordinary \emph{synchronous} function.
Calling \rcode{send} delivers the value and wakes the suspended receiver, which is exactly the obligation a reactor discharges.
We can see \rcode{send} as playing a reactor-like role, despite being a single function call rather than an event loop.

Two things set such a reactor apart from those of Section~\ref{sec:model}.
First, it discharges a finite, bounded collection of obligations rather than an unbounded stream: a one-shot channel has at most one receiver to wake.
Second, it runs a statically bounded number of times (one, in this example) and then terminates. Rust's ownership discipline consumes the sender, so \rcode{send} fires at most once per channel; after that single invocation the reactor is dead, and any obligation registered afterwards can never be discharged.
We model this as a \emph{terminating reactor} $\lkw{t-reactor}$ (Fig.~\ref{fig:lang-syntax-ext}), whose body $s$ runs to completion once per invocation, with no enclosing $\lkw{loop}$.
Its well-formedness predicate $\mathit{wfr}$ is that of a regular reactor (Section~\ref{sec:model-language}), minus the required outermost $\lkw{loop}$.

Verifying a $\lkw{t-reactor}$ reuses most of our existing design with only modest adaptation. $\VCR$ and the abstract domain are unchanged, with (R1) and (R2) of Definition~\ref{def:reactor-responsiveness} read \emph{per invocation} rather than per iteration; Reactor Soundness (Theorem~\ref{thm:reactor-soundness}) transports directly, since its proof already reasons about a single execution of the body.

Reactor Eventual Progression (Theorem~\ref{thm:reactor-eventual-progression}), by contrast, does \emph{not} translate as directly.
Recall that its proof relies on the reactor iterating infinitely often: an obligation is preserved across iterations by Definition~\ref{def:reactor-responsiveness} (R2) and woken by (R1) at some later iteration when its predicate holds.
A terminating reactor runs only once, so an obligation inserted after that single invocation is preserved by (R2) but never discharged.

Progression therefore holds only for obligations whose transfer \emph{strictly precedes} termination of the reactor.
The observation function $S$ in the current abstract state cannot witness this alone: a receiver that re-checks completion after transferring its obligation and one that does not both reach $\lPending$ with the same observation $S(p_c) = \afalse$.

Building on our first-class notion of monotonic predicates, we extend the abstract domain of Section~\ref{sec:verification-domain} with a \emph{precedes relation} ${\precedes} \subseteq \pred \times \pred$: a strict partial order representing observed temporal order between two monotonic predicates.
In particular, $p \precedes q$ means there \emph{exists} a moment with $p \in \hist$ and $q \notin \hist$.
This relation (and its transitive closure) can be efficiently maintained by our analysis.

We extend abstract states with an additional set $\precedes$ to capture all the observed precedes relations, \ie{} $\sigma = (S, \precedes, \srepos, \sloc)$.
The set $\precedes$ joins by intersection, keeping only orderings established on all incoming paths.

We additionally initialize $S$ with a dedicated monotonic predicate $\mathit{tfr}$ for ``the obligation has been transferred to the reactor''.
The \emph{strict obligation transfer (SOT)} condition is $\mathit{tfr} \precedes p_c$: the obligation must be transferred \emph{before} the completion predicate holds.

 The precedes relations can be tracked via two updated transfer rules, presented in Fig.~\ref{fig:strict-transfer} with no new annotation.
First, $\lkw{insert}(r, w, p)$ means the obligation has been transferred to a reactor; therefore, besides moving the obligation to $\srepos(r)$, we also set $S(\mathit{tfr}) = \atrue$.
Second, refining $S(p)$ to $\afalse$ (e.g., by entering the else-branch of an if-statement) extends $\precedes$ with $q \precedes p$ for every $q$ already observed to be true ($S(q) = \atrue$).

Let $\mathit{Term} \subseteq \Repo$ denote the set of repositories owned by terminating reactors. The pollable verification condition $\VCp$ is strengthened by a check on \emph{strict-obligation-transfer} that fires only when the pollable transfers its entry obligation into some $r \in \mathit{Term}$:
\begin{itemize}
    \item \text{(SOT$_{vc}$):} under $\sigma^{\top}_{\mathit{start}}$, at every reachable $\lkw{return}\;\lPending$ such that $(\hat{w}, \hat{p}, \_) \in \srepos(r)$ for some $r \in \mathit{Term}$: $\mathit{tfr} \precedes \hat{p}$.
\end{itemize}
A re-checking receiver that performs $\lkw{insert}$ (refining $S(\mathit{tfr})$ to $\atrue$) followed by a re-check (refining $S(p_c)$ to $\afalse$) now records the precedes relation $\mathit{tfr} \precedes p_c$, hence satisfying (SOT$_{vc}$); a non-re-checking receiver will fail (SOT$_{vc}$).

\begin{theorem}[Terminating Reactor Eventual Progression]
\label{thm:t-reactor-progression}
Let $r$ be the repository of a terminating reactor whose body $s$ satisfies $\VCR(s)$, and let $(w, p)$ be an obligation inserted into $r$ by a pollable satisfying \textnormal{SOT$_{\mathrm{vc}}$} for $r$. If the host invokes the reactor, then
\[ \square\bigl(\, p \in \hist \;\Rightarrow\; \Diamond\; w \in \wbuf \,\bigr). \]
\end{theorem}

\paragraph{Proof sketch.}
\textnormal{SOT$_{\mathrm{vc}}$} gives $\mathit{tfr} \precedes p$: the registration of $w$ into $r$ strictly precedes $p$ entering $\hist$.
The host's invocation establishes $p$ (it $\lkw{trigger}$s $p$, or observes it already $\atrue$) and, by $\VCR(s)$/(R1), $\lkw{take}$s and $\lkw{wake}$s every obligation in $r$ whose predicate holds.
Since $w$ was registered strictly before $p$, $(w,p)$ is present in $r$ at invocation time; (R1) fires $\lkw{wake}(w)$, and by the (wake) rule eventually $w \in \wbuf$.
A single invocation thus replaces the ``iterates infinitely often'' step of Theorem~\ref{thm:reactor-eventual-progression}.

The proof leaves one assumption uncovered, mirroring the regular reactor's reliance on its external predicates eventually holding: the host must eventually invoke the terminating reactor (e.g., call \rcode{send}). If a one-shot channel is never sent on, its receiver legitimately waits forever, a user-level failed obligation rather than a runtime fault.

\subsection{Extension 2: Fixpoint for Pollable Retry-loops}
\label{sec:fixpoint-retry-loops}

The base language requires pollable bodies to be loop-free ($\mathit{wfp}$, Section~\ref{sec:model-language}).
Sometimes, however, pollables themselves contend for bounded shared resources (a channel slot, a mutex permit), following a \emph{try-wait-retry} cycle: attempt to acquire the resource; on failure, $\lkw{insert}$ a waker and yield; on wake, retry. The reinvocation of pollables on wake only implies that the contended resource \emph{may} now be available, but success on the second call is not guaranteed.

Our modular analysis techniques do not support this pattern as described, which would require some kind of analysis around the implicit loop illustrated in \figref{flow}. However, a variant of this pattern which we have encountered in practice is to use an explicit \emph{try-wait-retry} loop in the pollable implementation itself; we describe here an extension which supports this case.

To support such retry-loops, we can admit $\lkw{loop}$ in pollable bodies and discharge it with a fixpoint; $\VCp$, both initial states, and the other transfer rules of Section~\ref{sec:verification} are unchanged.
The well-formedness condition $\mathit{wfp}$ is relaxed accordingly: $\lkw{loop}\;s$ is admitted in any statement position, and the body $s$ must contain \emph{at least} one $\lkw{return}$, so that the loop exit is reachable. Our analysis does \emph{not} attempt to check termination of this loop, which ultimately depends on an external predicate eventually holding; instead, this extension to pollable loops checks that \emph{provided} the loop eventually terminates, the necessary obligations will be satisfied.

Our fixpoint analysis converges because the abstract domain is a finite complete lattice and the transfer function is monotone.
For any pollable, the monotonic predicates, symbolic wakers, and repository names appearing in the pollable body are finite.
Therefore, the symbolic obligations $\sOb \subseteq \sWaker \times \pred \times \{\atrue, \top\}$ and the space of observation maps $S$ are also finite.
The state space $(\AbsState, \sqsubseteq, \sqcup, \bot_{\AbsState})$ is therefore a finite, complete lattice, with $\sqcup$ monotone in both arguments. Monotonicity of $\tfn[s]$ follows by structural induction on $s$: every rule of Fig.~\ref{fig:lang-transfer} is built from monotone operations on $\Obs$ and $\wp(\sOb)$.

The transfer function definitions extend with one new case for loops:
\begin{align*}
    \tfn[\lkw{loop}\;s](\sigma) &\;=\; \bot_{\AbsState}, \\
    h^{\ast} &\;\triangleq\; \mathrm{lfp}\bigl(\, h \,\mapsto\, \sigma \sqcup \tfn[s](h) \,\bigr).
\end{align*}
Here $h^{\ast}$ is the abstract state computed as the fixpoint above.
A $\lkw{loop}$ exits only through a $\lkw{return}$, so its result state is $\bot_{\AbsState}$; the return-site states that $\VCp$ inspects are gathered from inside $s$ during the evaluation of $\tfn[s](h^{\ast})$, exactly as the base analysis collects them for non-loop bodies, just with $h^{\ast}$ in place of the pollable's entry state. The body's transfer $\tfn[s](h)$ collects exactly the loop-back paths because any $\lkw{return}$ inside $s$ yields $\bot_{\AbsState}$, the identity of $\sqcup$.
The chain $h_0 = \bot_{\AbsState}$, $h_{n+1} = \sigma \sqcup \tfn[s](h_n)$ ascends and stabilizes in finitely many steps.
This generalizes to nested loops because monotonicity of the body's transfer is preserved by structural induction.

The proof of Pollable Soundness (Theorem~\ref{thm:pollable-soundness}) has a new case for the loop.
By induction on the iteration number, every concrete state at the loop head is abstracted by $h^{\ast}$: the entry state by $\sigma \sqsubseteq h^{\ast}$, and any future-iteration state, obtained by running the body once from a previous state, by $\tfn[s](h^{\ast}) \sqsubseteq h^{\ast}$.
Return-site observations collected under $h^{\ast}$ therefore over-approximate those of every concrete iteration, so $\VCp$ remains sound.
 
 \section{The \tool{} Tool and Evaluation}
\label{sec:impl-eval}

\begin{figure}[tbp]
    \centering
\includegraphics[width=\textwidth]{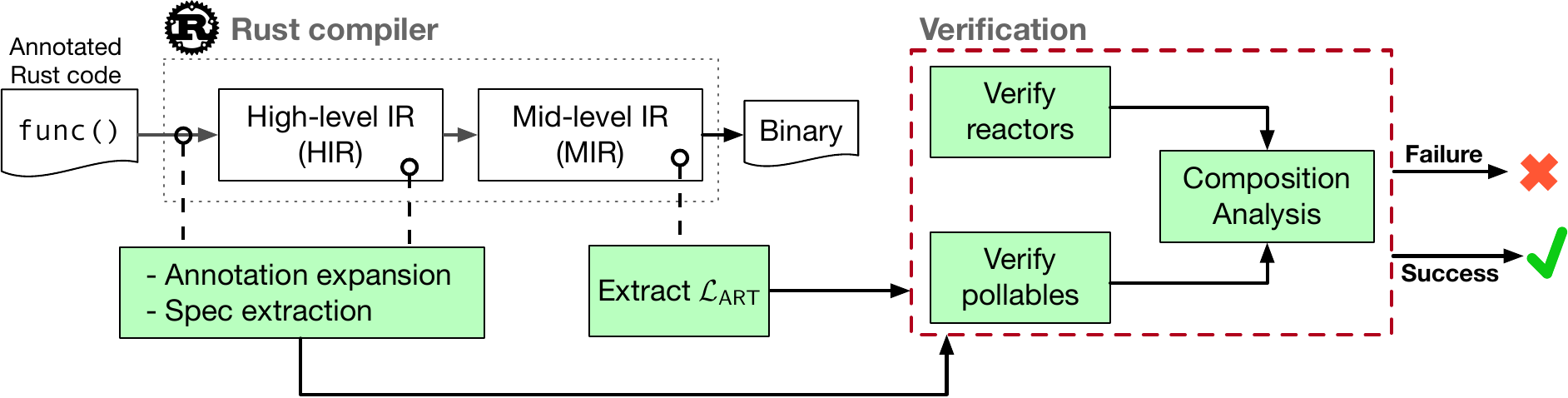}
    \caption{\tool design. The green boxes are part of \tool.}
    \label{fig:design}
\end{figure}

We implement our technique as a prototype, \tool{}, in roughly 5k lines of Rust. \tool{} is designed as a Rust compiler plugin~\cite{rustcplugin}; see Figure~\ref{fig:design}.
\tool{} enables modeling pollables/reactors of Rust async runtimes in $\lang$ via annotations and
 extracts an explicit $\lang$ model out of Rust's intermediate representation (Middle-level IR, MIR).
The extracted model can be manually inspected, and also automatically verified against $\VCp$ and $\VCR$ as described in Section~\ref{sec:verification}.
In Sec.~\ref{sec:eval} we use \tool{} to evaluate (1) the expressiveness of $\lang$ by modeling real-world pollables and reactors, and (2) the effectiveness of our static analyses.

Users model a pollable/reactor as an instance of $\lang$ through a macro library that overlays $\lang$ constructs on Rust code.
Macros expand to no-op markers in MIR for model extraction.
The annotated Rust code compiles and behaves identically to the original.
The macro syntax mirrors $\lang$: \rcode{mono\_pred!} declares a monotonic predicate; \rcode{obs!}, \rcode{obs\_tt!}, and \rcode{obs\_ff!} record observations on predicates; \rcode{trigger!}, \rcode{wake!}, \rcode{insert!}, and \rcode{take!} mark the corresponding effects; and \rcode{repo\_for!} marks iteration over a repository.
The entry points of pollables/reactors are tagged with \rcode{\#[jackdaw::pollable]} and \rcode{\#[jackdaw::reactor]}, and \rcode{bind!} binds the $\lang$ parameters of a component (its waker, completion predicate, and \public{} repositories) explicitly rather than having them inferred.

\tool{} traverses the MIR of annotated functions, captures the control-/data-flows and user annotations, and lifts the pollable/reactor implementations into $\lang$.
The lifting from MIR to $\lang$ is not formalized and contains heuristics (e.g., recovering nested branches and loops from a CFG).

The $\lang$ formalized in Sec.~\ref{sec:model} and the static analyses described in Sec.~\ref{sec:verification} are fully implemented (including the extensions).
We add an assume statement $\lkw{assume}(p, b)$ that refines a predicate $p$ to a specific value $b$ without branching, and a nondeterministic choice $\lkw{nd-match}$ for more straightforward translation from Rust.

\subsection{Evaluation}
\label{sec:eval}

We evaluate \tool{} on a representative subset of components drawn from three Rust async runtimes: smol~\cite{smol_docs}, futures~\cite{futures_docs}, and tokio~\cite{tokio_docs}. These span the three building blocks of async programs: scheduling (timers), I/O, and communication (one-shot and MPSC channels).
Tables~\ref{tab:pollables} and~\ref{tab:reactors} list every verified component with its source size (LoC, excluding blank lines and comments), total annotation cost (Anno LoC; Ratio = Anno / LoC), and per-component \emph{analysis time}, the cost of lifting to $\lang$ and discharging the verification conditions, measured as the mean of 300 runs.
Experiments ran on a 2.30~GHz Intel Core i9-9880H with 32~GB RAM under \rcode{rustc 1.91.0-nightly}.

\begin{table}[tbp]
\centering
\caption{Verification statistics for pollables.}
\label{tab:pollables}
\small
\setlength{\tabcolsep}{10pt}
\begin{tabular}{@{}l|ccccc@{}}
\toprule
\textbf{Pollable} & \textbf{Runtime} & \textbf{LoC} &
\textbf{\begin{tabular}[c]{@{}c@{}}Anno\\LoC\end{tabular}} &
\textbf{Ratio} &
\textbf{\begin{tabular}[c]{@{}c@{}}Analysis\\time ($\mu$s)\end{tabular}} \\
\midrule
EventListener        & smol    & 15 &  5 & 33\% &  24.6 \\
MPSC-send            & smol    & 28 &  6 & 21\% &  68.0 \\
MPSC-recv            & smol    & 27 &  6 & 22\% &  63.2 \\
MPSC-close           & smol    & 21 &  4 & 19\% &  39.0 \\
AsyncIO-Read/Write   & smol    & 29 &  6 & 21\% &  99.5 \\
AsyncIO-Ready        & smol    & 38 &  5 & 13\% &  86.2 \\
OneShot-recv         & futures & 23 &  8 & 35\% &  82.8 \\
OneShot-recv         & tokio   & 59 & 11 & 19\% & 152.2 \\
OneShot-closed       & tokio   & 45 & 10 & 22\% & 111.1 \\
\bottomrule
\end{tabular}
\end{table}

\begin{table}[tbp]
\centering
\caption{Verification statistics for reactors.}
\label{tab:reactors}
\small
\setlength{\tabcolsep}{6.5pt}
\begin{tabular}{@{}l|cccccc@{}}
\toprule
\textbf{Reactor} & \textbf{Runtime} & \textbf{Kind} & \textbf{LoC} &
\textbf{\begin{tabular}[c]{@{}c@{}}Anno\\LoC\end{tabular}} &
\textbf{Ratio} &
\textbf{\begin{tabular}[c]{@{}c@{}}Analysis\\time ($\mu$s)\end{tabular}} \\
\midrule
Timer / AsyncIO   & smol    & looping     & 59 & 11 & 19\% & 38.5 \\
OneShot-recv     & futures & terminating & 12 &  7 & 58\% & 10.9 \\
OneShot-recv     & tokio   & terminating & 12 &  7 & 58\% &  9.9 \\
OneShot-closed   & tokio   & terminating & 13 &  7 & 54\% & 10.6 \\
\bottomrule
\end{tabular}
\end{table}

\textbf{Effectiveness.}
For each verification condition (P1/P2/P3 of $\VCp$, R1/R2 of $\VCR$, and the strict-obligation-transfer condition from Section~\ref{sec:terminating-reactors}), we seed a violation in a real component and confirm \tool{} rejects it; no false negatives are observed across the suite.

\textbf{Specification burden.}
Each pollable verifies with 4 to 11 lines of annotation, giving an annotation-to-code ratio of 23\% on average.
Every pollable in the suite is specified with a \emph{single} monotone predicate.
Terminating reactors show a higher ratio because of their short implementations ($\le 13$ lines).

\textbf{Imprecision/Code modification.}
We encountered three cases where additional annotations or source code modifications are needed to overcome \tool{}'s imprecision or limitations.
\begin{enumerate*}
    \item Post-join re-assertion is needed when a completion decision is carried in a boolean flag across control-flow merges. This is a precision limitation of our abstract domain.
    \item Our model cannot handle tokio's fairness mechanism, which requires pollables to cooperatively yield. We use a special annotation in \tool{}, \rcode{ignore\_path!}, to prune such unsupported code paths.
    \item Tokio's OneShot-recv pollable contains paths requiring cross-invocation invariants to justify their returns; this requires additional ghost annotations to establish such invariants.
\end{enumerate*}

\textbf{Overhead.}
Per-component analysis is sub-millisecond throughout (10 to 152~$\mu$s; Tables~\ref{tab:pollables}~and~\ref{tab:reactors}).
Lifting dominates and scales with control-flow complexity; the heaviest case is tokio's one-shot receiver with its coop/\rcode{will\_wake}/state branches at $\approx$152~$\mu$s, while reactors are 10 to 40~$\mu$s.
These costs are negligible relative to Rust compilation, making \tool{} suitable for integrating more expensive analysis/verification techniques. \section{Limitations and Future Work}
\label{sec:limitations}

We discuss some of the current limitations and potential future work for our technique and its prototype implementation.
Known gaps in the existing model of $\lang$ include: a \emph{wake-one} primitive (non-deterministically discharge one obligation in a repository) for single-permit synchronization idioms (e.g., tokio's \rcode{Mutex}, \rcode{Semaphore}), and cross-poll reasoning that would let invariants established in a prior poll discharge later checks.

\textit{Sound extraction of $\lang$ models.}
Our soundness theorems are stated with respect to $\lang$, but real programs
are written in Rust. \tool{} bridges the two by lifting MIR to $\lang$ using a (currently trusted but fairly straightforward) translation of control flow, and, crucially, trusted user-supplied annotations. In particular, \tool{} does not check for mistakes in these annotations, such as triggering a monotonic event when the corresponding update has not been performed in the real memory.
We plan to investigate scalable approaches to close this gap or reduce the chance of errors in the user's modelling, by integrating other complementary static or dynamic analysis techniques.

\textit{Fine-grained concurrency.}
All operations in $\lang$, \eg{} \texttt{insert}, \texttt{take}, and \texttt{wake} are atomic operations under a sequential consistency model; our techniques, and especially the precedes relation extension defined in \secref{terminating-reactors} rely on sequential consistency. Real-world async runtime implementations such as \rcode{tokio} use weaker atomic memory accesses for high performance; so far, we have not seen examples in which extra observable behaviors could result, but our technique is not checking this formally when such primitives are used; adding such checks directly to our analyses in a way that matches the idioms used in this kind of code might be possible in the future.

\textit{Cancellation and task drop.}
Async tasks can typically be cancelled at any await point, at which time any wakers the task has registered with reactors must be reclaimed.
Our model currently has no notion of obligation revocation and no mechanism for a reactor to learn that a transferred obligation becomes irrelevant; adding this to our operational semantics would be straightforward. Although challenging in terms of general logic of user code, from the pollable and reactor perspectives we believe this extension should be reasonably simple, especially since wakers typically include logic to handle unnecessary extra calls.

\textit{Safety and functional correctness.}
Our analyses currently only check that obligations arisen from pollables managed by an async runtime are not lost, and we assume properties such as panic-freedom (and type-correctness, in $\lang$). It might be interesting in future to also be able to prove panic-freedom as well as to provide a formal basis for more-general verification of functional properties guaranteed by the runtime's mechanisms. We believe $\lang$ could be used as a basis for exploration of such techniques in the future.

\textit{Bounded responsiveness.}
Finally, eventual progression is a specific responsiveness property that only asserts progress in the arbitrarily-far future.
Users typically care about finer-grained responsiveness properties, \eg{} bounded latency, tail latencies, and fairness across tasks of different priorities.
Lifting our technique to stronger responsiveness properties (given some reasonable timing model) and/or trying to infer such guarantees would be interesting future work.
 \section{Related Work}
\label{sec:related}

Our work addresses the problem of verifying eventual progression, a liveness property ensuring that asynchronous runtimes make progress and do not indefinitely defer pending tasks. Some prior work has proposed approaches to reason about \emph{user code}~\cite{ganty2009verifying,kahlon2009async,ganty2012algasync,maiya2014androidraces} that relies on an async runtime. However, the underlying async runtime is typically considered part of the trusted computing base. Instead, our work treats errors in \emph{user code} (or other components such as the OS) as an orthogonal concern and focuses on critical runtime components that almost all user code implicitly relies upon.

\textbf{Compositional Verification of Responsiveness.}
Our work takes inspiration from Actor Services~\cite{summers2016actor}, which introduced a modular logic for verifying responsiveness and functional correctness of purely-actor-based programs. Actor Services employed a notion of logical obligations (reminiscent of ``causal obligations'' of Helm \etal{}~\cite{Helm'90}), to structure progress proofs under fair scheduling. Again, they address verification of \emph{user code}, while we focus on the more-specific but crucial concern of the responsiveness of Rust async runtime components.

We aim for similarly-modular reasoning to Actor Services, but our technique differs in several important aspects. We can verify implementation strategies that Actor Services cannot (in particular, reactor-like components which do not discharge obligations immediately, but which promise progress in a different fashion): reactors which handle multiple obligations in an interleaved fashion cannot be verified with this prior work. We also identify and build in monotonic events and their orderings as a first-class notion, which allows us to express subtle requirements that cannot be captured as responsiveness properties directly. We have also applied our technique to real-world code with \tool{}, enabling experimentation and evidence of potential impact in practice.

Anvil~\cite{sun2024anvil} similarly tackles liveness reasoning in Rust systems but in a very different setting. It provides a framework for verifying liveness properties of Kubernetes controllers using TLA specifications embedded within Verus~\cite{lattuada2024verus}. Anvil's ``Eventually Stable Reconciliation'' (ESR) property is conceptually related to our notion of Eventual Progression. Anvil requires users to encode their system logic manually in TLA and to prove ESR interactively within Verus. In contrast, we automatically verify liveness guarantees of async runtimes through lightweight specifications, rather than a user-driven proof framework.

Obligations have been used in other verification contexts: Leino \etal{}~\cite{LeinoMuellerSmans'10} proposed obligations as a first-class specification notion for enabling modular proofs of deadlock-freedom. This idea was extended by Bostr{\"o}m and M{\"u}ller~\cite{BostromMueller15} to enable verification of finite blocking for concurrent algorithms. These works are aimed at user code and do not directly apply to our setting; conversely, the simplifications afforded by our task-centric view or decomposition of responsibilities across pollables and reactors cannot be directly exploited for user code.

\textbf{Verification for Rust Programs.}
Rust's ownership model has motivated a growing body of verification work. RustBelt~\cite{jung2017rustbelt} provides a foundational semantics and verification technique based on Rust's type system, and deductive verifiers such as Prusti~\cite{prusti,astrauskas2022prusti}, Creusot~\cite{creusot}, and Verus~\cite{lattuada2024verus} support semi-automated functional correctness proofs. These techniques primarily target improvements to sequential reasoning made possible by Rust's guarantees; our key reasoning concepts of monotonic events and their orderings are novel. Of these, only Verus has been applied to the Rust async setting, by Lion~\cite{zhou2026lion}, discussed below.

Cutner and Yoshida~\cite{cutner2021asyncrust} formalize session-based async coordination of Rust user code that \emph{uses} async primitives such as channels. This helps prevent user code errors such as deadlocks; the responsibilities of the runtime are assumed in such work (as standard), while our work addresses these responsibilities directly.

\textbf{Verification of Liveness / Async Runtimes.}
To our knowledge, the first usage of monotonic predicates as a specific notion was in Hailpern and Owicki's pioneering work on modular specification and verification for liveness properties~\cite{hailpern1982modular}. This work presents a wide variety of specification idioms and shows why monotonic properties have useful properties at the boundaries between components. We take this idea further and make monotonic predicates into first class specification notions, along with the ability to specify and observe orderings of them. In addition, we show how to connect our reasoning to concrete program implementations.

Model checking has been used extensively to verify liveness properties~\cite{biere2002liveness}, sometimes using variations of rely-guarantee style proof techniques~\cite{jones1981development} to support modular verification~\cite{grumberg1994model}. While these techniques allow abstracting over the concrete interleavings of actions by other components, they do not provide the abstraction or simplicity of our task-centric view.

Liveness verification has been considered at the protocol level in the context of session types by Padovani \etal{}~\cite{Padovani'14}; this work does not connect to concrete implementations or support modular specification of separate components.

de Vilhena and Pottier develop a framework for separation logic reasoning for effect handlers~\cite{de2021separation}.
As part of the evaluation, they verified the safety property of an async runtime built with effect handlers. They do not consider liveness of async runtimes.

Independently of our work, Zhou \etal{}~\cite{zhou2026lion} developed \emph{Lion}, a clean-slate Rust async runtime with Tokio-compatible APIs, and mechanically verified its scheduling liveness in Verus~\cite{lattuada2024verus}.
Lion obtains verifiability by rebuilding the runtime, at roughly 40k lines of specification and proof for 3.4k lines of code, whereas \tool{} checks components of existing production runtimes as written, with lightweight annotations and automatic analyses.
Our decomposition from eventual progression into componentwise properties is proven once for every runtime expressible in $\lang$, whereas Lion composes module contracts per runtime through cross-module predicates.
Lion does \emph{not} verify the pollables that user code awaits.
Its end-to-end proof assumes that a task returning \lPending{} has registered a wake source.
Our verification considers properties of pollables, the condition (P1) to (P3) defined in Sec.~\ref{sec:model-properties} discharge exactly this assumption.
Conversely, Lion verifies its executor, which we trust (Section~\ref{sec:overview-ep}), and its proofs are machine-checked while our soundness argument is on paper.

 \section{Conclusion}
\label{sec:conclusion}

This work takes a first step toward verifying the \emph{liveness guarantees} of async runtimes. Whereas prior research has primarily focused on verifying the safety and correctness of async programs, we shift attention to the runtimes that are responsible for ensuring that tasks eventually complete.

We developed a detailed yet generic model of Rust async runtimes that captures how reactors, pollables, and schedulers cooperate to provide responsiveness.
We formalized this model in a core language $\lang$ and identified two component-level verification conditions whose conjunction soundly implies eventual progression of the runtime.
We automated these checks in \tool, a compiler extension for Rust that uses lightweight annotations and static analysis.
Our evaluation on \rcode{smol} and \rcode{tokio} async runtimes demonstrates that, with the two extensions of the core formalism, our approach verifies real pollable and reactor components written with various common idioms, while catching common responsiveness bugs.

Ultimately, our work advances the goal of \emph{verified asynchrony}: connecting the high-level liveness guarantees expected by developers with the concrete behavior of runtimes that implement them.
 
\clearpage

\bibliographystyle{splncs04}

\pagebreak
\appendix

\section{Well-formedness Conditions}
\label{app:wf}

This appendix gives the well-formedness predicates $\mathit{wfp}$ (for pollable bodies) and $\mathit{wfr}$ (for reactor bodies) referenced in Section~\ref{sec:model-language}.
They capture the programming paradigm not reflected in the bare syntax of $\lang$ (Fig.~\ref{fig:lang-syntax}).

\paragraph{Pollable bodies.}
A pollable $\pi$ is a statement satisfying the well-formedness predicate $\mathit{wfp}$: every execution path contains a $\lkw{return}\;v$ statement, and the statement contains no $\lkw{loop}$ or reactor-only constructs, i.e., $\lkw{take}$ and $\lkw{for}$.
Formally, $\mathit{wfp}(s) \triangleq \mathit{returns}(s) \,\wedge\, \mathit{pollable\text{-}only}(s)$, where:
\begin{align*}
    \mathit{returns}(\lkw{return}\; v) &= \top \\
    \mathit{returns}(s_1 \,;\, s_2) &= \mathit{returns}(s_1) \vee \mathit{returns}(s_2) \\
    \mathit{returns}(\lkw{if}\; p\; \lkw{then}\; s_1\; \lkw{else}\; s_2) &= \mathit{returns}(s_1) \wedge \mathit{returns}(s_2) \\
    \mathit{returns}(s) &= \bot \quad \text{otherwise}
\end{align*}
\begin{align*}
    \mathit{pollable\text{-}only}(\lkw{loop}\;s) = \mathit{pollable\text{-}only}(\lkw{take}(r, w))
        &= \bot \\
    \mathit{pollable\text{-}only}(\lkw{for}\;(w, p)\;\lkw{in}\;r\;\{\,s\,\}) &= \bot \\
    \mathit{pollable\text{-}only}(s_1 \,;\, s_2)
        &= \mathit{pollable\text{-}only}(s_1) \wedge \mathit{pollable\text{-}only}(s_2) \\
    \mathit{pollable\text{-}only}(\lkw{if}\; p\; \lkw{then}\; s_1\; \lkw{else}\; s_2)
        &= \mathit{pollable\text{-}only}(s_1) \wedge \mathit{pollable\text{-}only}(s_2) \\
    \mathit{pollable\text{-}only}(s) &= \top \quad \text{otherwise}
\end{align*}

\paragraph{Relaxed pollable bodies (with loops).}
\label{app:wfp-relaxed}
The loop extension admits $\lkw{loop}\;s$ in pollable bodies. The relaxed predicate $\mathit{wfp}'(s) \triangleq \mathit{returns}'(s) \,\wedge\, \mathit{pollable\text{-}only}'(s)$ replaces the $\lkw{loop}$ clauses of $\mathit{returns}$ and $\mathit{pollable\text{-}only}$ with
\begin{align*}
    \mathit{returns}'(\lkw{loop}\;s) &= \mathit{has\text{-}return}(s) \\
    \mathit{pollable\text{-}only}'(\lkw{loop}\;s) &= \mathit{pollable\text{-}only}'(s)
\end{align*}
where $\mathit{has\text{-}return}(s)$ asserts that $s$ contains a $\lkw{return}$ on at least one path:
\begin{align*}
    \mathit{has\text{-}return}(\lkw{return}\;v) &= \top \\
    \mathit{has\text{-}return}(s_1 \,;\, s_2)
        &= \mathit{has\text{-}return}(s_1) \vee \mathit{has\text{-}return}(s_2) \\
    \mathit{has\text{-}return}(\lkw{if}\; p\; \lkw{then}\; s_1\; \lkw{else}\; s_2)
        &= \mathit{has\text{-}return}(s_1) \vee \mathit{has\text{-}return}(s_2) \\
    \mathit{has\text{-}return}(\lkw{loop}\;s) &= \mathit{has\text{-}return}(s) \\
    \mathit{has\text{-}return}(s) &= \bot \quad \text{otherwise.}
\end{align*}
All other clauses of $\mathit{returns}$ and $\mathit{pollable\text{-}only}$ are unchanged. A $\lkw{loop}\;s$ is admitted iff its body uses no $\lkw{take}$ or $\lkw{for}$ and contains a reachable $\lkw{return}$, so loop exit is structurally possible.

\paragraph{Reactor bodies.}
A reactor body satisfies $\mathit{wfr}$: it never $\lkw{return}$s (reactors are intended to monitor the environment indefinitely) and contains no $\lkw{loop}$ outside the outermost $\lkw{loop}$ that surrounds it.
Formally:
\begin{align*}
    \mathit{wfr}(\lkw{skip}) = \mathit{wfr}(\lkw{trigger}(p)) = \mathit{wfr}(\lkw{wake}(w))
        &= \top \\
    \mathit{wfr}(\lkw{insert}(r, w, p)) = \mathit{wfr}(\lkw{take}(r, w))
        &= \top \\
    \mathit{wfr}(\lkw{return}\; v) = \mathit{wfr}(\lkw{loop}\; s)
        &= \bot \\
    \mathit{wfr}(s_1 \,;\, s_2)
        &= \mathit{wfr}(s_1) \wedge \mathit{wfr}(s_2) \\
    \mathit{wfr}(\lkw{if}\; p\; \lkw{then}\; s_1\; \lkw{else}\; s_2)
        &= \mathit{wfr}(s_1) \wedge \mathit{wfr}(s_2) \\
    \mathit{wfr}(\lkw{for}\; (w, p)\; \lkw{in}\; r\; \{\, s \,\})
        &= \mathit{wfr}(s)
\end{align*}

\section{Full Operational Semantics for $\lang$}
\label{app:opsem-full}

Figure~\ref{fig:lang-prog-step-full} gives the complete set of small-step program-semantics rules for $\lang$. The main-paper version (Fig.~\ref{fig:lang-prog-step}) omits some conventional rules; the full set is reproduced here for reference.

\begin{figure}[tbp]
    \centering
    \boxed{
    s \vbar \cfg \pstep s' \vbar \cfg'
    }
    \begin{mathpar}
    \inference{
    p \in \pint
    }{
    \lkw{trigger}(p) \vbar (\env, (\hist, \wbuf, \repos)) \pstep \lkw{skip} \vbar (\env, (\hist \cup \{p\}, \wbuf, \repos))
    }[(\text{trigger})]
    \and
    \inference{
    }{
    \lkw{wake}(w) \vbar (\env, (\hist, \wbuf, \repos)) \pstep \lkw{skip} \vbar (\env, (\hist, \wbuf \cup \{w\}, \repos))
    }[(\text{wake})]
    \and
    \inference{
    \repos' = \repos[r \mapsto \repos(r) \cup \{(w, p)\}]
    }{
    \lkw{insert}(r, w, p) \vbar (\env, (\hist, \wbuf, \repos)) \pstep \lkw{skip} \vbar (\env, (\hist, \wbuf, \repos'))
    }[(\text{insert})]
    \and
    \inference{
    (w, p) \in \repos(r) \\
    \repos' = \repos[r \mapsto \repos(r) \setminus \{(w, p)\}]
    }{
    \lkw{take}(r, w) \vbar (\env, (\hist, \wbuf, \repos)) \pstep \lkw{skip} \vbar (\env, (\hist, \wbuf, \repos'))
    }[(\text{take})]
    \and
    \inference{
    p \in \hist
    }{
    \lkw{if}\; p\; \lkw{then}\; s_1\; \lkw{else}\; s_2 \vbar \cfg \pstep s_1 \vbar \cfg
    }[(\text{if-t})]
    \and
    \inference{
    p \notin \hist
    }{
    \lkw{if}\; p\; \lkw{then}\; s_1\; \lkw{else}\; s_2 \vbar \cfg \pstep s_2 \vbar \cfg
    }[(\text{if-f})]
    \and
    \inference{
    \repos(r) = \{(w_1, p_1), \ldots, (w_k, p_k)\} \\
    s_i \triangleq s[w_i, p_i / w, p] \quad k \geq 1
    }{
    \lkw{for}\; (w, p)\; \lkw{in}\; r\; \{\, s \,\} \vbar (\env, \gst) \pstep
        s_1 \,;\, \cdots \,;\, s_k \vbar (\env, \gst)
    }[(\text{for})]
    \and
    \inference{
    \repos(r) = \emptyset
    }{
    \lkw{for}\; (w, p)\; \lkw{in}\; r\; \{\, s \,\} \vbar (\env, \gst) \pstep \lkw{skip} \vbar (\env, \gst)
    }[(\text{for-emp})]
    \and
    \inference{
    }{
    \lkw{loop}\; s \vbar \cfg \pstep s \,;\, \lkw{loop}\; s \vbar \cfg
    }[(\text{loop})]
    \and
    \inference{
    s_1 \vbar \cfg \pstep s_1' \vbar \cfg'
    }{
    s_1 \,;\, s_2 \vbar \cfg \pstep s_1' \,;\, s_2 \vbar \cfg'
    }[(\text{seq})]
    \and
    \inference{
    }{
    \lkw{skip} \,;\, s_2 \vbar \cfg \pstep s_2 \vbar \cfg
    }[(\text{seq-skip})]
    \and
    \inference{
    }{
    \lkw{return}\; v \,;\, s_2 \vbar \cfg \pstep \lkw{return}\; v \vbar \cfg
    }[(\text{seq-return})]
    \end{mathpar}
    \caption{Full small-step operational semantics for $\lang$.}
    \label{fig:lang-prog-step-full}
\end{figure} 
\end{document}